\documentclass[journal]{IEEEtran}
\usepackage{setspace}
\usepackage{cite}
\usepackage[pdftex]{graphicx}
\usepackage{amsmath}
\usepackage{amsfonts}
\usepackage[caption=false,font=footnotesize]{subfig}
\usepackage{xcolor}
\usepackage[T1]{fontenc}
\begin{document}
\title{Scenario Forecasting of Residential Load Profiles}

%\author{Ling~Zhang,~\IEEEmembership{Member,~IEEE,}
        %John~Doe,~\IEEEmembership{Fellow,~OSA,}
        %and~Jane~Doe,~\IEEEmembership{Life~Fellow,~IEEE}% <-this % stops a space

\author{Ling~Zhang,
        Baosen~Zhang% <-this % stops a space
\thanks{L. Zhang and B. Zhang are with the Department of Electrical and Computer Engineering, University of Washington, WA 98195 USA e-mail: lzhang18@uw.edu, zhangbao@uw.edu.}% <-this % stops a space
\thanks{L. Zhang is partially supported by the UW College of Engineering Fellowship.}}


% \markboth{Journal of \LaTeX\ Class Files,~Vol.~14, No.~8, August~2015}%
% {Shell \MakeLowercase{\textit{et al.}}: Bare Demo of IEEEtran.cls for IEEE Journals}

\maketitle

\begin{abstract}
Load forecasting is an integral part of power system operations and planning. Due to the increasing penetration of rooftop PV, electric vehicles and demand response applications, forecasting the load of individual and a small group of households is becoming increasingly important. Forecasting the load accurately, however, is considerable more difficult when only a few households are included. A way to mitigate this challenge is to provide a set of scenarios instead of one point forecast, so an operator or utility can consider a range of behaviors. This paper proposes a novel scenario forecasting approach for residential load using flow-based conditional generative models. Compared to existing scenario forecasting methods, our approach can generate scenarios that are not only able to infer possible future realizations of residential load from the observed historical data but also realistic enough to cover a wide range of behaviors.
Particularly, the flow-based models utilize a flow of reversible transformations to maximize the value of conditional density function of future load given the past observations. In order to better capture the complex temporal dependency of the forecasted future load on the condition, we extend the structure design for the reversible transformations in flow-based conditional generative models by strengthening the coupling between the output and the conditional input in the transformations.
The simulation results show the flow-based designs outperform existing methods in scenario forecasting for residential load by both providing more accurate and more diverse scenarios.

\end{abstract}

%\begin{IEEEkeywords}
%\del{Load forecasting, scenario generation, flow-based generative models, conditional models.}
%\end{IEEEkeywords}

% \fi
%\IEEEpeerreviewmaketitle

\section{Introduction}
%\input{intro.tex}
%% !TEX root=draft.tex

Distribution systems are becoming more dynamic and more decentralized because of the emergence of new technologies and services. Instead of operating distribution networks as passive systems, utilities start to account for distributed energy resources such as rooftop solar, electric vehicles and demand response programs. Since these resources are stochastic and intermittent, accurate forecasting of residential load for a single or a small number of households becomes important for operators to decide on whether to integrate distributed energy resources and where to deploy energy storage so as to match customers' demand and make better use of energy~\cite{Guo2012ManagePower}. In addition, accurate load forecasting on a small scale also allows customers to manage costs such as peak demand charges~\cite{Ji2018DataDrivenLM,Tascikaraoglu2014DSM}.

% For example, owing to the stochastic and intermittent characteristics of renewable energy resources, the increasing levels of renewables penetration is posing challenges to the operations of power systems. Under this circumstance, accurate forecasting of residential load on a small scale can be useful for operators to decide on whether to integrate distributed energy resources and where to deploy energy storage so as to match customers' demand and make better use of energy~\cite{Guo2012ManagePower}. In addition, accurate forecasting of residential load for an individual household also enables utilities to explore the flexibility in electricity consumption behaviors and have a better control of peak demand such that the expensive cost of peak load generators can be reduced~\cite{Ji2018DataDrivenLM},~\cite{Tascikaraoglu2014DSM}.

Compared to standard load forecasting used in transmission system operations, residential level forecasting have received less attention until relatively recently. For introductions and surveys on this topic the readers can refer to~\cite{Sevlian2018VaryLevels,veit2014household,Humeau13} and the references within. Despite these advances, residential load forecasting, especially for a single or a small number of households, remains a challenging problem for several reasons.
Firstly, individual load naturally exhibits higher volatility compared to a larger aggregation of loads because of the randomness of human behaviors and smaller base loads~\cite{Gajowniczek2017IndvForecast,Zhang2018SinglePerspect,Yildiz2018IndvForecast}. This makes achieving very accurate point forecasts fundamentally difficult and the standard metric of the distance between forecasted and realized values becomes less useful as a figure of merit~\cite{Hong2016Tutorial}.
Secondly, the increasing deployment of distributed solar, the popularity of plug-in hybrid electric vehicles (PHEVs) and the emerging trend of behind-the-meter energy storage bring even more uncertainties to the electricity demand of users.
Therefore, methodologies should capture and reflect these uncertainties in the forecasting process.
 % it becomes desirable to develop forecasting methods that can provide a group of possible scenarios in order to take into full consideration the uncertainties in the future load.}

An important method in load forecasting used to describe the future uncertain associated with a load is \textit{scenario forecasting}~\cite{Hyndman2014Forecasting}. Different than the conventional deterministic point forecasting approach \cite{Hippert2001NN,Chen2004Competition,Weron2006StatForecast}, which generates the most likely forecast for the future load as a single estimate, scenario forecasting provides a wide range of possible realizations of the future that can occur. Scenario forecasting can be more useful compared to probabilistic forecasts by not only informing operators of the uncertainties about the future in the form of prediction intervals or quantile forecasts, but also generating a set of plausible time series for early planning~\cite{Hong2016Tutorial}. This also opens more possibilities of generating realistic residential load profiles to compensate for the lack of measured residential load datasets. These datasets are difficult to collect because of changes in occupancy and spotty deployment of smart meters \cite{Pflugradt2017SynthLoad,Narayan2018ConstructLoad}.
% Furthermore, in recent years, as data privacy gains importance \cite{Rouf2012NeighborWatch}, utilities are being more restrictive on giving access to residential load dataset.
As a result of insufficient measured data, average load profiles are commonly used in research studies, which may lead to misleading results due to a lack of diversity~\cite{Linssen2017LoadInfluence}.
In these settings, scenario forecasting provides a promising method to generating artificial residential load profiles that have similar properties to measured data and hence can be used in downstream tasks in power systems.

Previous works on generating scenario forecasts for residential load falls into two main categories. The first category is to make use of the point load forecasts coming from the pre-trained models and then add noise to them~\cite{Xie2017NormalAssump}. Specifically, the residuals of the point forecasts are modeled with a normal distribution, which is then added back to the original point load forecasts to generate a set of possible scenarios. The other category of methods take advantage of the relationship between the weather and the load, and generate probabilistic forecasts of load based on simulated weather scenarios~\cite{McSharry2005PeakPLF,Fan2012ShortTerm,Dordonnat2016GEFCom,Hong2018PLF}. For example, a group of weather scenarios are created based on the historical data, and then each generated weather scenario is fed into the point load forecasting model to obtain a different set of point forecasts for the load.
The former category suffers from the fact it creates scenarios centered at the point forecast, which may not capture the diversity in load behaviors, especially if there are multiple modes in the data. The latter category can generate more diverse scenarios, but does not overcome the fundamental issues since it pushes the question to that of how to select a set of good weather profiles. More fundamentally, both methods are based on point forecasts, which are designed to be the deterministic solutions that minimize a distance metric. However, the goal of scenario forecasting is different, being that we want to generate i.i.d. samples of possible future load realizations.

% However, both categories of methods are built on point forecasts of the load, and may not be sufficient to provide a comprehensive picture of future events. On one hand, point forecasting models are always trained to produce the most likely forecasts, as a result, the scenarios generated based on them may lack diversity and fail to capture possible extremes in the future load. On the other hand, the point load forecasting model with a single set of parameters may not be able to identify the complex dynamics existing in the residential load especially when the load data show significant volatility.

Recently, generative models based on (deep) neural networks have been applied to scenario forecasts generation to overcome the challenges in more traditional methods. The works in \cite{Chen2018ModelFreeRS,Chen2018Unsupervised} use the Generative Adversarial Network (GAN) \cite{Goodfellow2014GenerativeAN} to generate scenarios for renewable resources (i.e., wind and solar). This method is then extended by \cite{Gu2018GANbasedRL} to generate scenario forecasts for residential load. Particularly, the work in \cite{Gu2018GANbasedRL} built the GAN model based on the Auxiliary Classifier GAN (ACGAN) \cite{Odena2017ConditionalIS} to generate load profiles with specific load patterns. It is worth noting that these generative models are not really forecasting models, since they can only include discrete valued conditional informations (e.g., winter vs. summer days). However, in most practical forecasting applications, the side information to be conditioned on is typically continuous-valued, like the past observations of the residential load.
% in the sense that they do not take the realized values into account, whereas most forecasting algorithms uses observed values to predict the future. This common for most most generative models, where any conditioning on side information can only take discrete values.

% The models in \cite{Chen2018ModelFreeRS,Gu2018GANbasedRL} can be made to use conditional information, but in both papers are conditioned on discrete-valued event labels to control the type of scenarios to be generated. However,

% In \cite{Chen2018Unsupervised}, the authors combine the GAN model with the point forecasting method, and generate the renewable scenarios based on the past observations in two steps. Specifically, they first found out the corresponding latent space of GAN through optimization methods such that the generated scenarios from this latent space could be consistent with the past observations and also not conflict with the given point forecasts. Then they sample points from the determined latent space and provide them as the inputs to the pre-trained GAN for scenario generation.

In this paper, we focus on directly providing scenario forecasts for future residential load without the help of point forecasts. Furthermore, we are interested in generating scenarios by conditioning on the historical realizations of the residential load.
It is interesting to note that the performance of the conditional GAN (CGAN) \cite{Mirza2014ConditionalGAN} is not satisfactory for this task because the conditional information is continuous and vector valued, and the training of GAN models is notoriously unstable because of its two constantly competing components-the generator and the discriminator \cite{Arjovsky2017WGAN}. Thus, we adopt flow-based generative models \cite{Dinh2014NICE,Dinh2017RealNVP,Kingma2018GlowGF} for this task.

\begin{figure}[!t]
\centering
\includegraphics[width=4in]{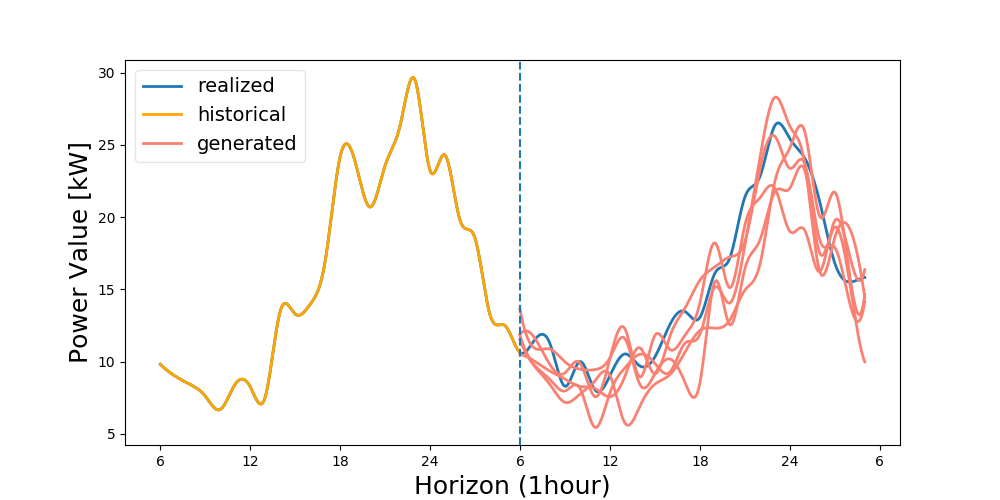}
\caption{
Realized residential load versus a group of generated scenario forecasts using our methods. Red curves are generated scenarios, the blue curve is the realized load data and the black curve is the historical load data. Generated scenario forecasts exhibit both accuracy and diversity.
 }
\label{intro:examples}
\end{figure}

Different than the GAN model, which bypasses modeling data distribution in its objective function, the flow-based generative models directly maximize the value of the probability density function (PDF) of the data and employ a series of specially-designed reversible transformations to enable efficient computation.
%\todo{again, say a few words of how flow based model does this}
Our experiment results show that the flow-based generative models achieve significantly better performance in generating high-quality residential load scenario forecasts given the past observations. Fig. \ref{intro:examples} shows examples of the generated daily scenarios using our methods by conditioning on the previous day's realized load.

The main contributions of this paper are:
\begin{enumerate}
\item We incorporate continuous-valued side information into the training process of flow-based generative models to make the model consider the correlation between the data and the condition. We also extend the structure design in the reversible transformations of flow-based models by enhancing the dependency of the output on the conditional input for better conditional scenario forecasting.
%\todo{instead saying we propose two flow designs, can we say we extend the flow method by doing... to get a new structure?} 
%We have also proposed two designs for the normalizing flow structure in the flow-based conditional generative models such that the complex temporal dependencies between the input data and the conditional data could be captured. 
%In addition, we provide some theoretical insights into how to reduce the variance of the generated scenarios by trading off the training objective of flow-based generative models against the Wasserstein distance between the learned data distribution and the true one.
\item By training flow-based conditional generative models, we propose a novel approach for forecasting residential load scenarios from given historical observations. Compared to existing methods in scenario forecasting for residential load, our methods can generate scenarios that not only provide more accurate forecasts for future load evolution but also cover a wider range of possibilities of future realizations.

\item We also provide some theoretical insights into how to reduce the variance of generated scenarios by controlling the trade-off between the original training objective of flow-based models and the newly added Wasserstein distance metric that indicates the distance between the modeled distribution and the true one. 
Simulation results show, by adding a weighted Wasserstein distance metric to the training objective, generated scenarios can become more centered around the realized load.

%Simulation results show the scenarios generated by our methods are not only realistic-looking but also able to extrapolate different possible realizations of the future load based on the given information. We show that our methods outperform the prevalent methods for residential load scenario forecasts.

%\item[3)] We also provide theoretical insights into the trade-off between only optimizing the training objective of flow-based generative models and optimizing both the training objective
\end{enumerate}

All of the code and data described in this paper are publicly available at~https://github.com/zhhhling/June2019.git.
The rest of the paper is organized as follows: Section \ref{flow-based-generative-models} introduces the flow-based generative models; Section \ref{conditional-scenario-generation} extends the flow-based models to conditional generative models, and employ the flow-based conditional generative models to conditional scenario forecasting for residential load; simulation results are illustrated and evaluated in Section \ref{simulation-results}; and Section \ref{conclusions} concludes the paper.

\section{Flow-Based Generative Models}
\label{flow-based-generative-models}
%\input{flow.tex}
%% !TEX root=draft.tex
In this section, we introduce the flow-based generative models \cite{Dinh2014NICE,Kingma2018GlowGF,Dinh2017RealNVP}.
Unlike the other types of generative models, such as Generative Adversarial Networks (GANs) \cite{Goodfellow2014GenerativeAN} and Variational Encoders (VAEs) \cite{Kingma2014AutoEncodingVB}, the training objectives of which either avoid constructing the PDF of data or utilize a lower bound instead, flow-based generative models are trained to directly maximize the value of modeled PDF.
We first review the change-of-variable technique and use this technique to formulate the objective function for training flow-based models, then we talk about the architectures adopted in flow-based models that enable the efficient computation of training objective.

Consider a $D-$dimensional data variable $X$, and $\mathbf{x}$ is one realization of it. Denote the true value of the PDF of $X$ at the point $\mathbf{x}$ by $p_X(\mathbf{x})$, and we train a flow-based generative model to estimate this value. Specifically, we first draw a $D-$dimensional latent variable $\mathbf{z}$ from a simple prior distribution $p_Z$ and provide it as the input to the flow-based model. Suppose there exists a bijective mapping $f:\mathbb{R}^D\longrightarrow \mathbb{R}$ such that $f(\mathbf{x}) = \mathbf{z}$ and $f^{-1}(\mathbf{z}) = \mathbf{x}$. Then, according to the change-of-variable formula, the density function of $X$ at the given point $\mathbf{x}$ can be represented by \cite{Dinh2017RealNVP}
\begin{align}
& p_X(\mathbf{x}) = p_Z(f(\mathbf{x}))|\mbox{det}(\frac{\partial f(\mathbf{x})}{\partial \mathbf{x}^T})|
\label{pdf}\\
& \log p_X(\mathbf{x}) = \log p_Z(f(\mathbf{x})) + \log |\mbox{det}(\frac{\partial f(\mathbf{x})}{\partial \mathbf{x}^T})|.\label{log-pdf}
\end{align}
Typically, the density function $p_Z$ is chosen to be standard multivariate Gaussian, i.e., $\mathcal{N}(0, \mathbf{I})$, and $\mbox{det}(\frac{\partial f(\mathbf{x})}{\partial \mathbf{x}^T})$ is the determinant of the Jacobian matrix of $f$ at $\mathbf{x}$. 
Since the ground truth bijective mapping $f$ that can map the distribution $p_Z$ to the true PDF of $X$ is unknown, we implement a parameterized bijective function $f_{\mathbf{\theta}}$ that can be learned by training the flow-based model. From the change-of-variable formula in (\ref{pdf}) or (\ref{log-pdf}), the modeled PDF of $X$ under the mapping of $f_{\mathbf{\theta}}$ is given by
\begin{align}
& p_X(\mathbf{x};\mathbf{\theta}) = p_Z(f_{\mathbf{\theta}}(\mathbf{x})) |\mbox{det}(\frac{\partial f_{\mathbf{\theta}}(\mathbf{x})}{\partial \mathbf{x}^T})| \label{likelihood}\\
& \log p_X(\mathbf{x};\mathbf{\theta}) = \log p_Z(f_{\mathbf{\theta}}(\mathbf{x})) + 
\log |\mbox{det}(\frac{\partial f_{\mathbf{\theta}}(\mathbf{x})}{\partial \mathbf{x}^T})|\label{log-likelihood}
\end{align}
where the modeled PDF of $X$ $p_{X}(\mathbf{x};\theta)$ can be considered as a function of the parameter $\mathbf{\theta}$, which is called the likelihood function and denoted by $L(\mathbf{\theta})$; the log of $L(\mathbf{\theta})$ is called the log-likelihood function, and denoted by $l(\mathbf{\theta})$. Using the maximum likelihood estimation (MLE) method, we can  train the flow-based model to choose the appropriate value of $\mathbf{\theta}$ such that the likelihood in (\ref{likelihood}) or the log-likelihood in (\ref{log-likelihood}) is maximized:
\begin{align}
\max_{\mathbf{\theta}}   \log p_X(\mathbf{x};\mathbf{\theta}) = \log p_Z(f_{\mathbf{\theta}}(\mathbf{x})) +
\log|\mbox{det}(\frac{\partial f_{\mathbf{\theta}}(\mathbf{x})}{\partial \mathbf{x}^T})|. \label{MLE}
\end{align}

In flow-based models, the parameterized bijective function $f_{\mathbf{\theta}}$ is chosen to be the composition of a sequence of reversible transformations, that is, $f_{\mathbf{\theta}} = f_1 \circ f_2 \circ \cdots \circ f_K $, such that the mapping from $\mathbf{x}$ to $\mathbf{z}$ and the inverse mapping from $\mathbf{z}$ to $\mathbf{x}$ can be represented as follows:
\begin{align}
&\mathbf{x}\overset{f_1}{\longrightarrow}\mathbf{h}_1
\overset{f_2}{\longrightarrow} \mathbf{h}_2 \cdots
\mathbf{h}_{K-1}\overset{f_K}{\longrightarrow} \mathbf{z}\label{forward}\\
&\mathbf{z}\overset{f_K^{-1}}{\longrightarrow}\mathbf{h}_{K-1}
\overset{f_{K-1}^{-1}}{\longrightarrow} \mathbf{h}_{K-2} \cdots
\mathbf{h}_1\overset{f_1^{-1}}{\longrightarrow} \mathbf{x}.\label{inverse}
\end{align}
The sequence of reversible transformations in (\ref{forward}) is called a normalizing flow \cite{Kingma2018GlowGF}. 
Based on the design of the normalizing flow in (\ref{forward}), the log-determinant of the Jacobian matrix $\frac{\partial f_{\mathbf{\theta}}(x)}{\partial x^T}$ can be written as follows by using the chain rule:
\begin{align}
\log |\mbox{det}(\frac{\partial f_{\mathbf{\theta}}(x)}{\partial x^T})| = \sum_{i=1}^{K} \log |\mbox{det}(\frac{\partial \mathbf{h}_i}{\partial \mathbf{h}_{i-1}^T})|\label{Jabobian}
\end{align}
where $\mathbf{h}_0 = \mathbf{x}$, and $\mathbf{h}_K = \mathbf{z}$. 
To facilitate the computation of the log-determinants of Jacobian in (\ref{Jabobian}), each reversible transformation $f_i$ in (\ref{forward}) and (\ref{inverse}) is implemented as an affine coupling layer. Take the affine coupling layer design in Real-valued Non-Volume Preserving  (RealNVP) model \cite{Dinh2017RealNVP} as an example.
Given a $D-$dimensional input $\mathbf{x}$, we can split it into two parts, $\mathbf{x}_{1:d}$ and $\mathbf{x}_{d+1:D}$, where $d<D$. Then the output $\mathbf{y}$ of an affine coupling layer is given by
\begin{align}
&\mathbf{y}_{1:d} = \mathbf{x}_{1:d}\label{coupling-layer-1}\\
&\mathbf{y}_{d+1:D} = \mathbf{x}_{d+1:D} \odot \exp(s(\mathbf{x}_{1:d}))
+ t(\mathbf{x}_{1:d})\label{coupling-layer-2}
\end{align}
where $\odot$ represents element-wise product, and $s(\cdot)$ and $t(\cdot)$ are scaling and translating functions, respectively.
\footnote{
Note that the other flow-based generative model, Non-linear Independent Component Estimation (NICE) model \cite{Dinh2014NICE}, uses a similar affine coupling layer structure as in (\ref{coupling-layer-1}) and (\ref{coupling-layer-2}) but without scaling, and the latest generative flow (Glow) model \cite{Kingma2018GlowGF} adopts the same affine coupling layer as RealNVP.
}
Following the transformations in (\ref{coupling-layer-1}) and (\ref{coupling-layer-2}), the Jacobian matrix of $\mathbf{y}$ at $\mathbf{x}$ is a lower-triangular matrix
\begin{align}
\frac{\partial \mathbf{y}}{\partial \mathbf{x}^T} =
\begin{bmatrix}
\mathbf{I}_d & 0 \\
* & \mbox{diag}(\exp(s(\mathbf{x}_{1:d})))
\end{bmatrix}\label{J-of-coupling-layer}
\end{align}
and its log-determinant is simply $\mbox{sum}(s(\mathbf{x}_{1:d}))$. Note that, in (\ref{J-of-coupling-layer}), the operation $\mbox{diag}(\cdot)$ constructs a diagonal matrix from a vector, and the value of $*$ has no impact on the log-determinant of this Jacobian.

In order to get the series of transformations in (\ref{forward}), multiple coupling layers like (\ref{coupling-layer-1}) and (\ref{coupling-layer-2}) are combined in an alternating way to construct a normalizing flow\cite{Dinh2017RealNVP}. As a result, the log-determinant of the Jacobian matrix $\frac{\partial f_{\mathbf{\theta}}(x)}{\partial x^T}$ in (\ref{log-likelihood}) is just a sum of lower-triangular matrices' log-determinants, which makes the efficient computation of the training objective in (\ref{MLE}) possible.

\section{Conditional Scenario Generation}
\label{conditional-scenario-generation}
%\input{conditional.tex}
%% !TEX root=draft.tex
In this section, we first show flow-based generative models can be extended to conditional generative models by providing some side information $\mathbf{c}$ as the conditional input in the training process.
Particularly, different than previous conditional generative models which typically condition on discrete-valued categorical labels~\cite{Mirza2014ConditionalGAN}, flow-based conditional generative models can condition on continuous-valued data, such as time series observations.
Aside from the basic structure of flow-based conditional generative models, we also provide a new structure design for the transformations in the normalizing flow in order to capture as much information as possible from the conditional input.
At the end of this section, we describe how to employ flow-based conditional generative models to scenario forecasting for residential load by considering the historical observations.

\subsection{Conditional Flows}
\label{conditional-flows}
Considering the data sample $\mathbf{x}\in\mathbb{R}^{D}$ and the associated side information $\mathbf{c}\in\mathbb{R}^{D'}$, we train a flow-based generative model to estimate the value of the conditional PDF of $X$ at the point $\mathbf{x}$ given $\mathbf{c}$, i.e., $p_{X|C}(\mathbf{x}|\mathbf{c})$. Specifically, we first draw a latent variable $\mathbf{z}$ from distribution $p_Z$. Then we construct a parameterized bijective function $f_{\mathbf{\theta}}$ that takes $\mathbf{c}$ as an extra input such that $f_{\mathbf{\theta}}(\mathbf{x};\mathbf{c}) = \mathbf{z}$ and $f_{\mathbf{\theta}}^{-1}(\mathbf{z};\mathbf{c}) = \mathbf{x}$. Using the change-of-variable formula in (\ref{pdf}) or (\ref{log-pdf}), the modeled conditional PDF of $X$ under the mapping of $f_{\mathbf{\theta}}$ can be written as
\begin{align}
&p_{X|C}(\mathbf{x}|\mathbf{c}) = p_Z(f_{\mathbf{\theta}}(\mathbf{x};\mathbf{c}))
|\mbox{det}(\frac{\partial f_{\mathbf{\theta}}(\mathbf{x};\mathbf{c})}{\partial \mathbf{x}^T})|
\label{cond-pdf}\\
&\log p_{X|C}(\mathbf{x}|\mathbf{c}) = \log p_Z(f_{\mathbf{\theta}}(\mathbf{x};\mathbf{c}))
+ \log|\mbox{det}(\frac{\partial f_{\mathbf{\theta}}(\mathbf{x};\mathbf{c})}{\partial \mathbf{x}^T})|.
\label{log-cond-pdf}
\end{align}
Using the MLE method, we train the flow-based model to maximize the conditional likelihood of $X$ in (\ref{cond-pdf}) or the conditional log-likelihood of $X$ in (\ref{log-cond-pdf}):
\begin{align}
\max_{\mathbf{\theta}}   \log p_{X|C}(\mathbf{x}|\mathbf{c};\mathbf{\theta}) = \log p_Z(f_{\mathbf{\theta}}(\mathbf{x};\mathbf{c})) +
\log|\mbox{det}(\frac{\partial f_{\mathbf{\theta}}(\mathbf{x};\mathbf{c})}{\partial \mathbf{x}^T})|. \label{cond-MLE}
\end{align}
Suppose we collect $N$ independent identically distributed (i.i.d.) samples $(\mathbf{x}_1, \mathbf{c}_1),\cdots,(\mathbf{x}_N, \mathbf{c}_N)$ from the ground-truth conditional distribution $p_{X|C}(\mathbf{x}|\mathbf{c})$, following the objective function in  (\ref{cond-MLE}), we can train the flow-based model on this dataset through the following optimization:
\begin{align}
\max_{\theta} \sum_{i=1}^N\log p_{X|C}(\mathbf{x}_i|\mathbf{c}_i) & =  \sum_{i=1}^N\log p_Z(f_{\mathbf{\theta}}(\mathbf{x}_i;\mathbf{c}_i)) \\
& + \sum_{i=1}^N\log|\mbox{det}(\frac{\partial f_{\mathbf{\theta}}(\mathbf{x}_i;\mathbf{c}_i)}{\partial \mathbf{x}_i^T})|.
\label{cond-objective}
\end{align}

%Denote the parameters associated with the bijective function $f$ by $\mathbf{\theta}$, then the resulting objective function for training flow-based conditional generative models can be written as
%\begin{align}
%\max_{\theta} \sum_{i=1}^N\log p_{X|C}(\mathbf{x}_i|\mathbf{c}_i) = \sum_{i=1}^N\log p_Z(f_{\theta}(\mathbf{x}_i;\mathbf{c}_i)) +
%\sum_{i=1}^N\log|\mbox{det}(\frac{\partial f_{\theta}(\mathbf{x}_i;\mathbf{c}_i)}{\partial \mathbf{x}_i^T})|\label{cond-objective}
%\end{align}
%where $(\mathbf{x}_1, \mathbf{c}_1),\cdots,(\mathbf{x}_N, \mathbf{c}_N)$ are $N$ independent identically distributed (i.i.d.) samples taken from the ground-truth conditional distribution $p_{X|C}(\cdot|\cdot)$, and the distribution $p_Z$ is chosen to be standard multivariate Gaussian.
When constructing the parameterized bijective function $f_{\mathbf{\theta}}$ for flow-based conditional generative models, we provide $\mathbf{c}$ as an extra input to both the scaling and translating functions for each affine coupling layer in the normalizing flow, that is, given the input $\mathbf{x}$, the output of the affine coupling layer is given by
\begin{align}
&\mathbf{y}_{1:d} = \mathbf{x}_{1:d}\label{vanilla-1}\\
&\mathbf{y}_{d+1:D} = \mathbf{x}_{d+1:D} \odot \exp(s(\mathbf{x}_{1:d}, \mathbf{c}))
+ t(\mathbf{x}_{1:d}, \mathbf{c}).
\label{vanilla-2}
\end{align}
We call the flow-based conditional generative models with the basic structure in (\ref{vanilla-1}) and (\ref{vanilla-2}) the \textit{vanilla}-flow.

We can see from (\ref{vanilla-1}) and (\ref{vanilla-2}), in each affine coupling layer of \textit{vanilla}-flow, only one part of the output is affected by the condition. In order for the output to have more dependencies on the conditional input, we extend the structure design in \textit{vanilla}-flow to get a new structure. Specifically, the part of the input that remains unchanged in (\ref{vanilla-1}) also goes through scaling and translating transformations to reach the output:
\begin{align}
&\mathbf{y}_{1:d} = \mathbf{x}_{1:d} \odot \exp(s(\mathbf{c}))
+ t(\mathbf{c}) \label{reinforced-1}\\
&\mathbf{y}_{d+1:D} = \mathbf{x}_{d+1:D} \odot \exp(s(\mathbf{x}_{1:d}, \mathbf{c}))
+ t(\mathbf{x}_{1:d}, \mathbf{c})\label{reinforced-2}.
\end{align}
We call the flow-based conditional generative models with this new structure the \textit{reinforced}-flow.
It is worth pointing out that, in (\ref{reinforced-1}), the scaling and translating functions associated with $\mathbf{x}_{1:d}$ only have the condition $\mathbf{c}$ as their inputs.
With this design, the Jacobian matrix of the transformation given in (\ref{reinforced-1}) and (\ref{reinforced-2}) is still lower-triangular:
\begin{align}
\frac{\partial \mathbf{y}}{\partial \mathbf{x}^T} =
\begin{bmatrix}
\mbox{diag}(\exp(s(\mathbf{c})))  & 0 \\
* & \mbox{diag}(\exp(s(\mathbf{x}_{1:d}, \mathbf{c})))
\end{bmatrix}.
\label{J-reinforced}
\end{align}
As a result, the calculation of the log-determinants of Jacobian and hence the training objective in (\ref{cond-MLE}) for \textit{reinforced}-flow is as efficient as that for \textit{vanilla}-flow.

\begin{figure}[ht]
\centering
\subfloat[Training Phase]{\includegraphics[width=3in]{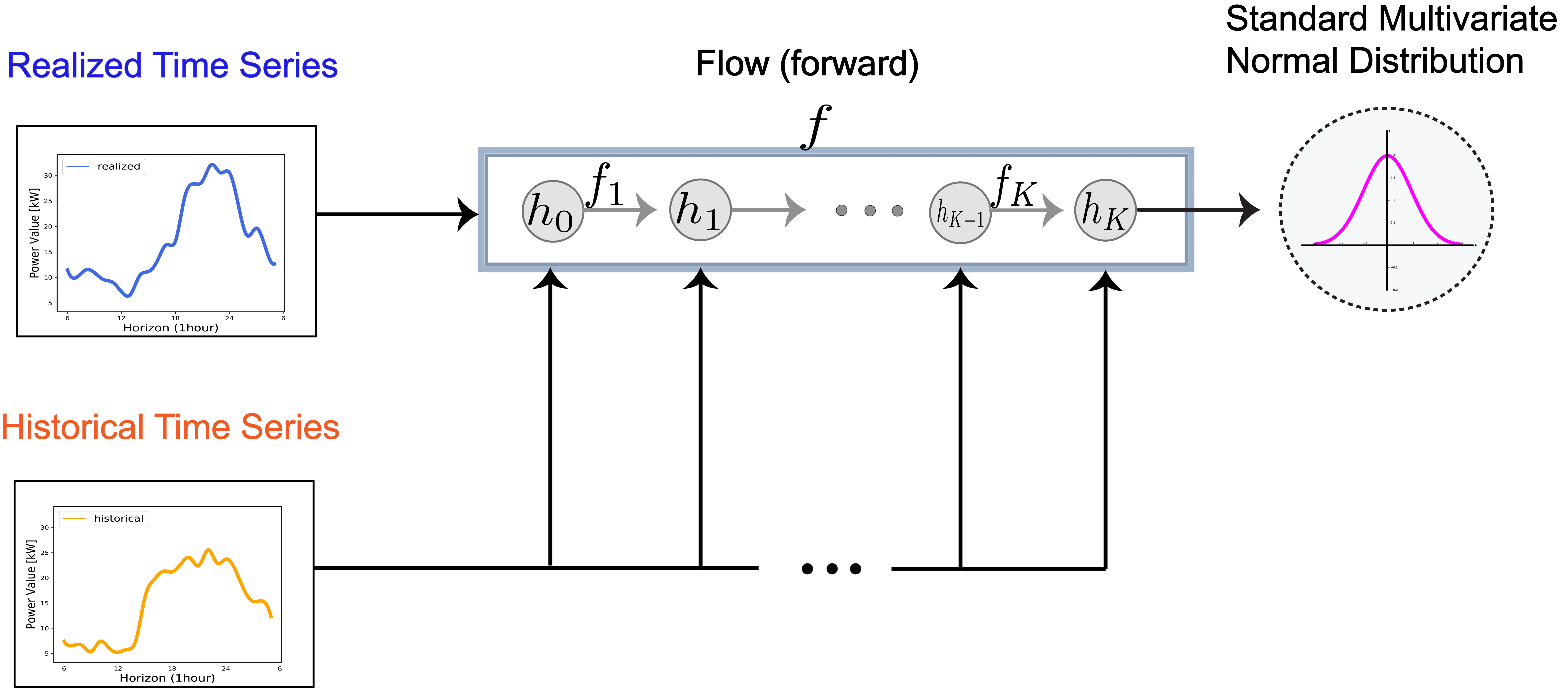}
\label{training}}
\hfil
\subfloat[Forecasting Phase]{\includegraphics[width=3in]{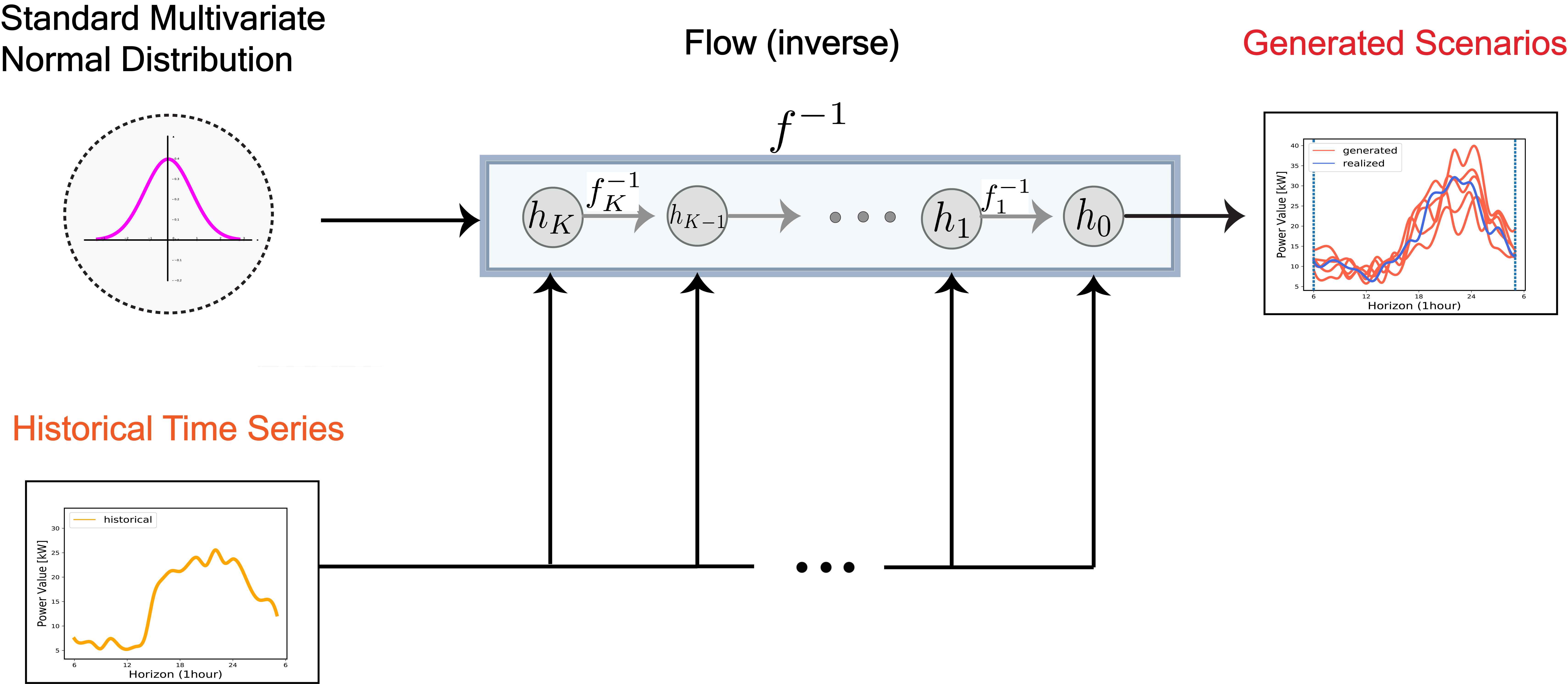}
\label{generation}}
\caption{
The architecture of the flow-based model that we use for training (a) and residential load scenario forecasting (b). In the training phase, we provide the realized load time series as the data input and the historical time series as the conditional input to the flow-based model, which learn the bijective function that can map the modeled data distribution to the standard multivariate normal distribution by maximizing the conditional likelihood of the training set. After training is completed, given a historical time series as conditional input, the trained flow-based model takes a handful of samples that come from the Gaussian distribution as data inputs, and uses the learned bijective function to transform these noise samples to a group of scenario forecasts that follow the same distribution as the realized time series.
}
\label{procedure}
\end{figure}
Now we apply flow-based conditional generative models to scenario forecasting for residential load. Specifically, we focus on generating a set of scenario forecasts for future load from the given historical realizations. Assuming at time $T$, we have $h$ observations of the previous realized residential load, which are collected in the vector $\mathbf{q}_{past} = \{q_{T-h}, \cdots, q_{T-1}\} \in\mathbb{R}^{h}$. Given this historical data $\mathbf{q}_{past}$, we generate scenario forecasts for future $k$ time points, and $k$ is referred to as the forecasting horizon. Suppose we have access to the realized load over the $k$ look-ahead time points, which is denoted by $\mathbf{q}_{true}\in\mathbb{R}^{k}$, then $(\mathbf{q}_{past}, \mathbf{q}_{true})$ can constitute a training sample for training flow-based models, with $\mathbf{q}_{past}$ as the conditional input and $\mathbf{q}_{true}$ as the data input. Note that the data sample $(\mathbf{q}_{past}, \mathbf{q}_{true})$ is actually a time series of length $(h+k)$. If we have a long time series of the realized residential load over $L$ time points and $L\gg h+k$, then we can break down this long time series into small pieces where each piece corresponds to a time-series data sample and overlaps with the following one. To be specific, the $i-$th piece consists of $(\mathbf{q}_{past}^{i}, \mathbf{q}_{true}^{i})$, and the following $(i+1)-$th piece consists of $(\mathbf{q}_{past}^{i+1}, \mathbf{q}_{true}^{i+1})$, where $\mathbf{q}_{past}^{i+1}$ is exactly $\mathbf{q}_{true}^{i}$. In this way, we can construct our training dataset $\{(\mathbf{q}_{past}^{i}, \mathbf{q}_{true}^{i})\}_{i=1}^N$ for training flow-based models through the optimization in (\ref{cond-objective}).

%The input to the model at the $i-$th data point is $\mathbf{q}_{true}^{i}$, and the condition is $\mathbf{q}_{past}^{i}$.
%The training objective is to maximize the empirical conditional log-likelihood of the
%future $k$ time points' residential load given the past $h$ observations
%training set in order to find the optimal bijective transformation that can map from the standard multivariate Gaussian distribution to our target one.
Suppose the learned value of parameter $\mathbf{\theta}$ through the optimization in (\ref{cond-objective}) is denoted by $\hat{\mathbf{\theta}}$, and the associated learned bijective function is denoted by $f_{\hat{\mathbf{\theta}}}$. Given any available historical residential load time series $\mathbf{q}_{past}$ of length $h$, we can generate scenario forecasts for the following $k$ time points using the inverse function of $f_{\hat{\mathbf{\theta}}}$ as follows
\begin{align}
\hat{\mathbf{q}}_{future} = f_{\hat{\mathbf{\theta}}}^{-1}(\mathbf{z}; \mathbf{p}_{past}) \label{conditional-genration}
\end{align}
where $\mathbf{z}\in\mathbb{R}^k$ is any sample taken from the standard multivariate Gaussian distribution. If we can take as many as $m$ samples from the standard multivariate Gaussian distribution, then we can produce $m$ conditional scenarios for the future residential load through (\ref{conditional-genration}). The architecture of the flow-based model that we use for training and residential load scenario forecasting is given in Fig. \ref{procedure}.

\subsection{Flows with Wasserstein Distance}
\label{W-flows}
From Section \ref{flow-based-generative-models} we know the objective function for training flow-based generative models is to maximize the log-likelihood of the training data. It turns out this objective is equivalent to minimizing the Kullback-Leibler (KL) divergence between the true data distribution and the modeled one~\cite{Murphy2012MLP}. To show this, recall the objective function in (\ref{MLE}), and rewrite it as follows
\begin{align}
\hat{\mathbf{\theta}} =
\arg\max_{\mathbf{\theta}}
\sum_{i=1}^N \log p_X(\mathbf{x}_i|\mathbf{\theta}).\label{theta_hat}
\end{align}
where $\hat{\mathbf{\theta}}$ is the set of trained parameters.
Suppose the ground-truth value of the parameter $\mathbf{\theta}$ is denoted by $\mathbf{\theta}^{\star}$. Since the optimization problem in (\ref{theta_hat}) is independent of the ground truth value $\mathbf{\theta}^{\star}$, we can rewrite it as follows
\begin{align}
\hat{\mathbf{\theta}} & =
\arg\max_{\mathbf{\theta}}
\frac{1}{N}\sum_{i=1}^N \log p_X(\mathbf{x}_i|\mathbf{\theta})\\
& = \arg\max_{\mathbf{\theta}}
\frac{1}{N}\sum_{i=1}^N \log p_X(\mathbf{x}_i|\mathbf{\theta}) -
\frac{1}{N}\sum_{i=1}^N  \log p_X(\mathbf{x}_i|\mathbf{\theta}^{\star})\\
& = \arg\max_{\mathbf{\theta}}
\frac{1}{N} \sum_{i=1}^N \log \frac{p_X(\mathbf{x}_i|\mathbf{\theta})}{p_X(\mathbf{x}_i|\mathbf{\theta}^{\star})}\\
& = \arg\min_{\mathbf{\theta}}
\frac{1}{N} \sum_{i=1}^N \log \frac{p_X(\mathbf{x}_i|\mathbf{\theta}^{\star})}{p_X(\mathbf{x}_i|\mathbf{\theta})},
\label{kl-empirical}
\end{align}
The expression in (\ref{kl-empirical}) is exactly the empirical KL divergence between the true data distribution $p_X(\mathbf{x}|\mathbf{\theta}^{\star})$ and the modeled data distribution $p_X(\mathbf{x}|\mathbf{\theta})$. That is to say, the goal of the flow-based generative models is to minimize the KL divergence between the true data distribution and the modeled one.

There are well known advantages and disadvantages in using the KL divergence as the distance measure between two probability distributions~\cite{durrett2019probability}. An undesirable effect of using it in scenario generation is that it is asymmetric: in the cases where, for a given point $\mathbf{x}$, the true data distribution $p_X(\mathbf{x}|\mathbf{\theta}^{\star})$ is significantly non-zero, while the learned distribution $p_X(\mathbf{x}|\mathbf{\theta})$ is close to zero, then the KL divergence can be infinitely large; however, when the true data distribution $p_X(\mathbf{x}|\mathbf{\theta}^{\star})$ is close to zero at the point $\mathbf{x}$, but the learned distribution $p_X(\mathbf{x}|\mathbf{\theta})$ is significantly non-zero, the KL divergence is small. As a result, by minimizing the KL divergence between the learned data distribution and the true one, the learned data distribution tend to spread out, which leads to good coverage of the data points that come from the true data distribution, but also tend to create superfluous data points that are not a part of the true distribution. Notably, the learned data distribution can have a relatively larger variance than the true one.

\begin{figure}[ht]
\centering
\includegraphics[width=3.5in]{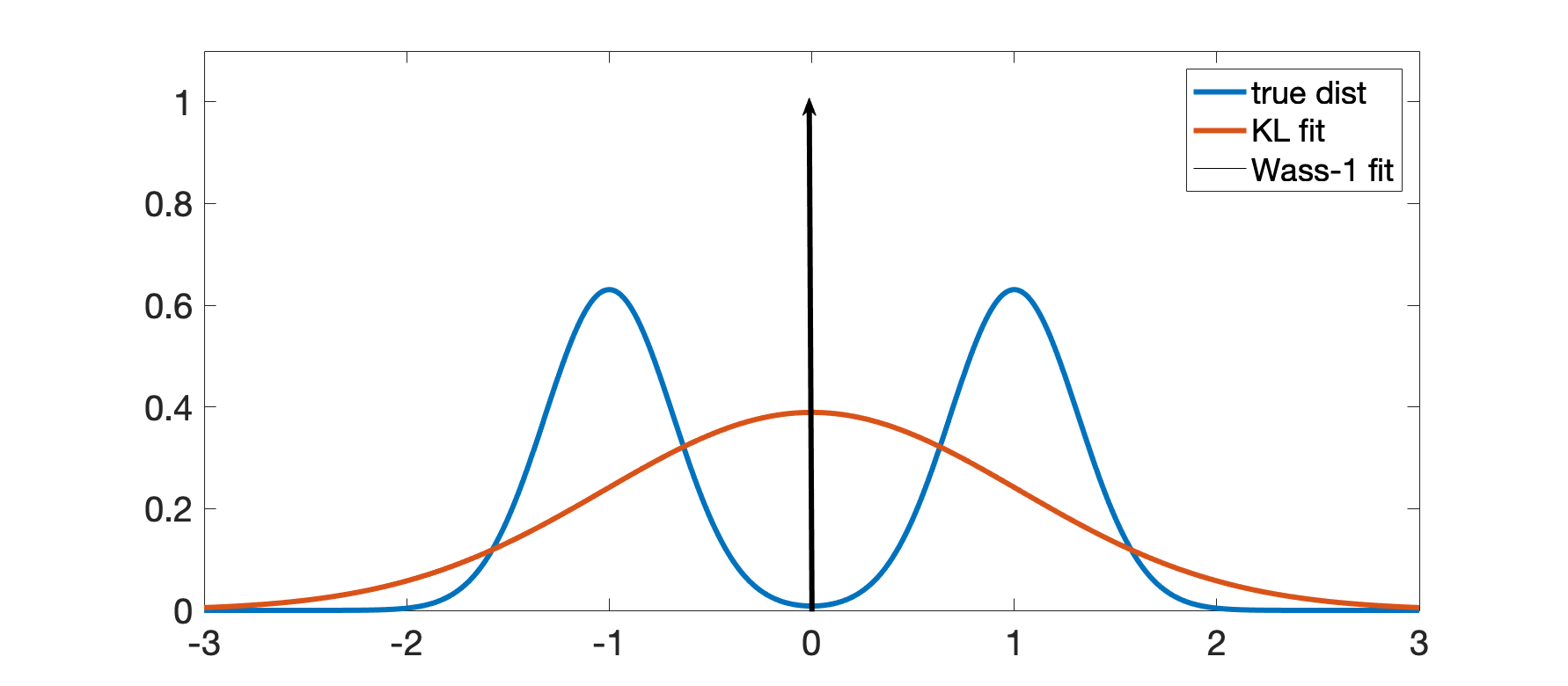}
\caption{An example of fitting a mixed Gaussian distribution with a zero mean Gaussian distribution using the KL divergence and Wasserstein distance. Under the KL divergence, the fitted distribution tend to be spread out to cover the true distribution, while under the Wasserstein distance the fitted distribution tend to concentrate to minimize the distance between the two components.}
\label{trade-off}
\end{figure}

A simple example of this effect can be found in Fig.~\ref{trade-off}. Suppose there is a one-dimensional Gaussian mixture model with two equally weighted components. The first component has mean $\mu_1 = -1.0$ and variance $\sigma_0^2 = 0.1$, while the second has mean $\mu_2 = 1.0$ and variance $\sigma_0^2 = 0.1$.
Now suppose we hope to fit a zero-mean Gaussian distribution $Q = \mathcal{N}(0, \sigma^2)$ to this Gaussian mixture model by tuning values of the variance $\sigma^2$ to minimize the KL divergence.
This is a one dimensional minimization problem and can be solved by simply graphing the KL divergence while varying $\sigma^2$. It turns out the optimal $\sigma^2$ is around $1.05$ and the ground-truth distribution $p_X(x)$ and the optimal distribution $Q^{KL} = \mathcal{N}(0, 1.05)$ are shown in Fig. \ref{trade-off}. This figure shows that the fitted Gaussian distribution covers both components of the Gaussian mixture model and even has a much larger variance.
% \begin{figure*}[!t]
% \centering
% \subfloat[Fitting by minimizing the KL divergence metric]{\includegraphics[width=3.5in]{KL-Divergence.png}
% \label{fig:KL-Divergence}}
% \hfil
% \subfloat[Fitting by minimizing the Wasserstein distance metric]{\includegraphics[width=3.5in]{KL-Divergence.png}
% \label{fig:Wasserstein}}
% \caption{An example of fitting a mixed Gaussian distribution with a zero mean Gaussian distribution using the KL divergence and Wasserstein distance. Under the KL divergence, the fitted distribution tend to be spread out to cover the true distribution, while under the Wasserstein distance the fitted distribution tend to concentrate to minimize the distance between the two components.}
% \label{trade-off}
% \end{figure*}

Since in many situations the parametrized distribution we learn by minimizing the KL divergence would not include the true data distribution, it becomes desirable to balance the impact of KL divergence using another distance metric. To this end, we use the Wasserstein distance as a regularizer in the flow objective. This is inspired by the performance of the  Wasserstein based generative adversarial networks (WGAN) \cite{Arjovsky2017WGAN}. The impact of the Wasserstein distance can be seen again in~\ref{trade-off}. Here we fit the Gaussian mixture model by minimizing the Wasserstein distance between the ground-truth distribution $p_X(x)$ and the fitted distribution $Q = \mathcal{N}(0, \sigma^2)$ in order to decide the optimal value of $\sigma$. It turns out that there is a closed-form expression for the Wasserstein-$1$ distance between two one-dimensional distributions~\cite{ruschendorf1985wasserstein}:
\begin{align}
W_1(P,Q) = \int_0^1 |F^{-1}(z) - G^{-1}(z)|dz
\label{eq:Wasserstein}
\end{align}
where $F$ and $G$ are the cumulative distribution functions (CDFs) of $P$ and $Q$. Using~\eqref{eq:Wasserstein}, the optimal $\sigma$ turns out to be to $0$, which is the delta function (point distribution) at $0$ as shown in Fig.~\ref{trade-off}. This fit is much more ``concentrated'' than the fit using KL divergence, but it do not cover the true distribution.

The intuition gained in this toy example carries over to more complex and higher dimensional distributions, which leads to a natural solution of combining the  objective of the flow-based generative models (KL divergence) with the Wasserstein distance. This both decreases the variance of the generated scenarios and at the same time generate plausible future realizations that have significantly non-zero probabilities to occur. Specifically, we add a weighted Wasserstein distance metric to the training objective of flow-based generative models:
%The combined objective function can be represented as
\begin{align}
\max_{\mathbf{\theta}} p_X(\mathbf{x}|\mathbf{\theta}) + \beta W(p_X(\mathbf{x}|\theta^{\star}, p_X(\mathbf{x}|\theta))
\label{combined-obj}
\end{align}
We call the flow-based generative models that are trained using the combined objective function in (\ref{combined-obj}) $\mathcal{W}$-flows, and the flow-based conditional generative models trained in this way are  called \textit{conditional} $\mathcal{W}$-flows, particularly, which include the \textit{vanilla}-$\mathcal{W}$-flow and the \textit{reinforced}-$\mathcal{W}$-flow.

\section{Simulation Studies}
\label{simulation-results}
%\input{results.tex}
%% !TEX root=draft.tex
In this section, we study the performance of our proposed flow-based approach in conditional scenario forecasting for residential load. We focus on generating daily scenario forecasts by conditioning on observed realizations in the previous day at an hourly resolution.
We first show that our approach can provide more accurate forecasts for daily residential load in the form of quantile  forecasts and prediction intervals compared to the standard scenario forecasting method of adding noise to point forecasts. Particularly, we use Gaussian noise in this paper. We also show generated scenarios are visually similar to the realized residential load. Also scenario forecasting do not generate a central forecast, we use the median of a group of generated scenarios for illustration purposes. Then we quantitatively evaluate our approach to show the generated scenarios using our approach outperform those generated from the standard method by having enhanced reliability and sharpness.

The experiments in this paper are implemented using Pytorch \cite{Paszke2017AutomaticDI} and the latest Glow model \cite{Kingma2018GlowGF}.
The flow-based model adopted in the simulation is composed of chained $9$ blocks of reversible transformations.
Except that the scaling and translating functions that only take the condition as inputs are implemented as fully-connected Neural Netoworks (NNs), all the other scaling and translating functions in the normalizing flow of the adopted flow-based model are implemented using three-layer 1D-Convolutional Neural Networks (1D-CNNs).
Batch-normalization is applied in 1D-CNNs before every layer except the input layer to increase the stability of learning. The activation functions for all scaling transformations are tanh function while rectified linear units (ReLUs) are used as the activation functions in all translating transformations. All models in this paper are trained using Adam optimizer \cite{Kingma2015Adam}. All data and codes can be found at~https://github.com/zhhhling/June2019.git.

\subsection{Dataset Description}

All experiments in this paper use the residential load dataset from Dataport, which is a database hosted by the Pecan Street Corporation~\cite{web:pecanstreet}. Hourly household power consumption from $1/1/2013$ to $12/31/2017$ in Austin, Texas, are used. This dataset consist of $128$ households, but only $105$ of them are consistent in the dataset. Therefore, we restrict our dataset to these $105$ households. Particularly, the data from $1/1/2013$ to $10/1/2017$ is used fore training and validation, and the last three month's data is used for testing. We conduct our simulation experiments for a varying number of aggregated households, and, particularly, the residential load from single household, $10$ households and $100$ households are used. At each aggregation level, we repeat the experiment for $10$ independent runs.

\subsection{Simulation Results}
\label{qualitative-evaluation}
\begin{figure}[ht]
\centering
\subfloat[Single household]{\includegraphics[width=3in]{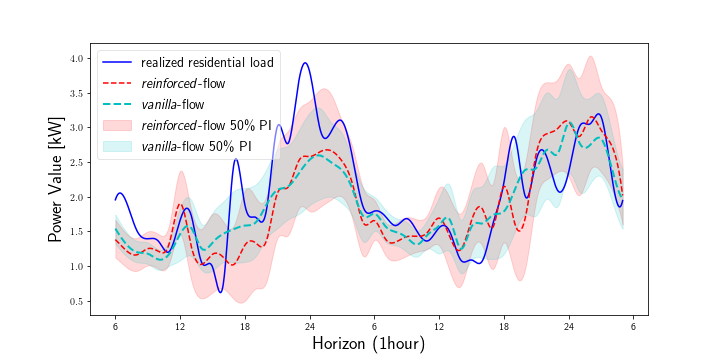}
\label{level-1}}
\hfil
\subfloat[10 households]{\includegraphics[width=3in]{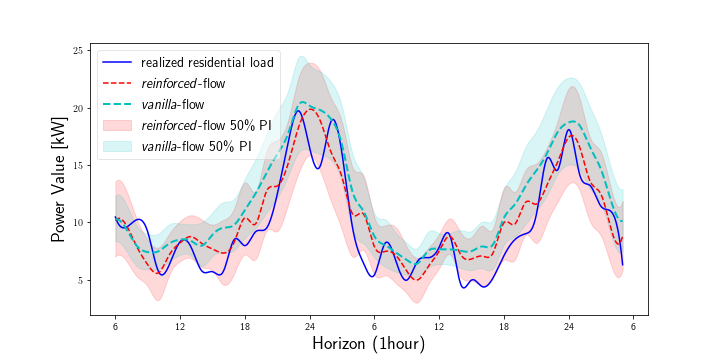}
\label{level-10}}
\hfil
\subfloat[100 households]{\includegraphics[width=3in]{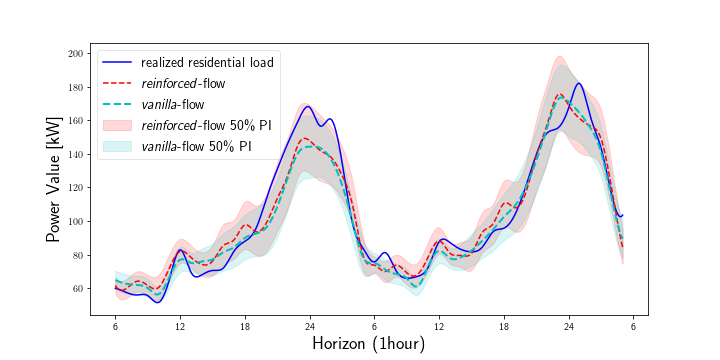}
\label{level-100}}
\caption{
Generated scenario forecasts for residential load from single household (a), 10 households (b), and 100 households (c) using two structures of the flow-based model: \textit{reinforced}-flow (red) and \textit{vanilla}-flow (cyan). The blue curve is the realized residential load, dashed curves represent the medians of the generated scenarios, and colored banded areas indicate the $50\%$ prediction interval of generated scenario forecasts.
Although the scenarios generated by both flow structures generally follow the realized load, reinforced-flow has better performance since its 50\% prediction interval covers more of the realized load.}
\label{all-levels}
\end{figure}

We first show the scenario forecasting results of our flow-based approach using the two structures, i.e., \textit{reinforced}-flow and \textit{vanilla}-flow, for a varying number of aggregated households. Specifically, we select one 48-hour long sample from the test data of each aggregation level. All the samples are taken from the same month in a year, i.e., October, 2017. These samples are given in Fig. \ref{all-levels}. In each sample, we plot the median of the generated scenarios versus the realized load for illustrative purposes. We also plot the $50\%$ prediction interval (PI), i.e., the interval between the $25^{\mbox{th}}$ and the $75^{\mbox{th}}$ quantile, as a colored belt to show the confidence in the generated scenario forecasts.

We can see from Fig. \ref{all-levels} that the median of the generated scenarios using both structures can give accurate day-ahead forecasts for the coming time and power value of the load peak and load valley, although median forecasts are less accurate for the residential load from  single household in Fig. \ref{level-1} than for $100$ households in Fig. \ref{level-100}. Particularly, the median of the generated scenarios is comparable to the realized load in visual quality, especially when more than one household are included.
In addition, it is worth noting from Fig. \ref{level-100} that the realized residential load of 100 households is completely covered by the $50\%$ prediction interval of the generated scenarios using both structures, and the width of the $50\%$ prediction interval in Fig. \ref{level-100} is also the narrowest among all three samples.
When only single household is considered, as shown in Fig. \ref{level-1}, the $50\%$ prediction interval of the generated scenarios using \textit{reinforced}-flow can cover more parts of the realized load in comparison to those using \textit{vanilla}-flow. This is because the residential load from single household has large randomness and is typically hard to be forecasted accurately. Under this circumstance, \textit{reinforced}-flow can have better performance than \textit{vanilla}-flow due to the improved structure design of the affine coupling layer in \textit{reinforced}-flow, as discussed in Section \ref{conditional-flows}, such that there is a stronger coupling between the future load and the historical realizations that are used for conditioning.
For the residential load of 10 households, although the randomness is reduced by aggregation, the realized load is still corrupted by a certain degree of noise and show volatility. In this setting, we can see from Fig. \ref{level-10} that \textit{reinforced}-flow again shows better performance than \textit{vanilla}-flow in the coverage of the actual load realization using the 50\% prediction interval.
\footnote{
Since the residential load from single household is notoriously difficult to be forecasted accurately because of the large randomness, and the forecasting for an aggregation of 100 households is typically a easier task  \cite{Sevlian2018VaryLevels}, we only use the forecasting results for 10 households as examples in the remaining part of Section \ref{simulation-results} to illustrate the performance of our approach. For the complete data and code, please refer to~https://github.com/zhhhling/June2019.git.}

\begin{figure*}[ht]
\centering
\includegraphics[width=5in]{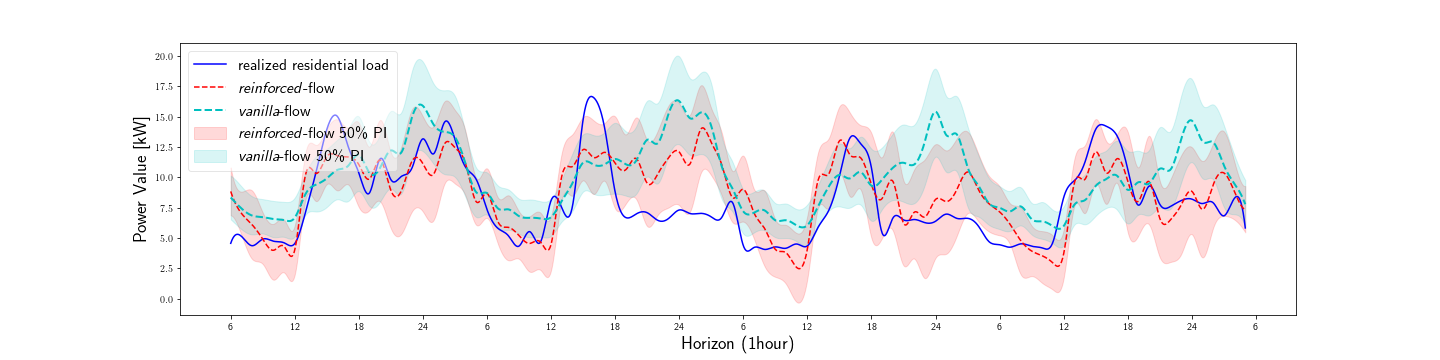}
\caption{
An example to show the better performance of \textit{reinforced}-flow (red) than \textit{vanilla}-flow (cyan) in generating scenarios that are more closely related to the pattern in the historical observations.
The blue curve is the realized residential load, the dashed curves represent the medians of the generated scenarios, and the colored banded areas indicate the $50\%$ prediction interval.
The scenarios generated by \textit{reinforced}-flow are more correlated to the historical realizations and can provide more accurate forecasts even when the realized load goes through sudden changes.
}
\label{compare-two-flows}
\end{figure*}
Recall from Section \ref{conditional-flows} that we extend the structure design in \textit{vanilla}-flow by strengthening the coupling of the condition to the output in the affine layers to get the new structure in \textit{reinforced}-flow. In order to validate if the new structure in \textit{reinforced}-flow can have better performance in learning more rich information from the condition, we show a series of 4 days' scenario forecasting results using both structure designs in Fig. \ref{compare-two-flows} and plot the medians and the $50\%$ prediction intervals of the generated scenarios. We can see from Fig. \ref{compare-two-flows} that there is a change in the pattern of the realized load from the second day with the second peak around $24:00$ becoming flat. On the third day, the median of the generated scenarios using \textit{reinforced}-flow also goes through a similar change and become even more accurate in forecasting the realized load evolution on the fourth day. The $50\%$ prediction interval of the generated scenarios using \textit{reinforced}-flow also experience similar changes as those in the historical realization and the median, and surround the realized load closely. By contrast, the median and the $50\%$ prediction interval of the generated scenarios using
\textit{vanilla}-flow have little difference from day to day and tend to disregard the changes in the condition.

\begin{figure}[ht]
\centering
\subfloat[Autoregressive and \textit{vanilla}-flow]{\includegraphics[width=3in]{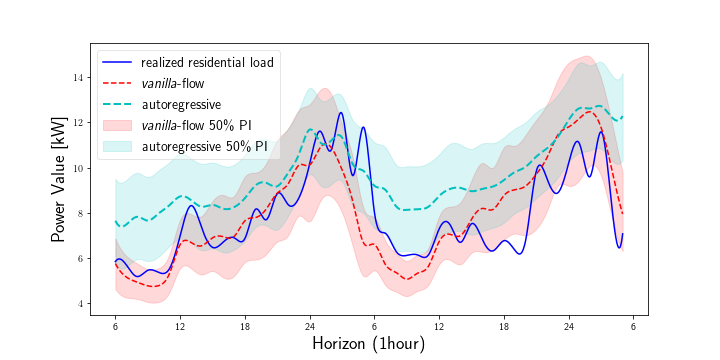}
\label{AR-vanilla}}
\hfil
\subfloat[Autoregressive and \textit{reinforced}-flow]{\includegraphics[width=3in]{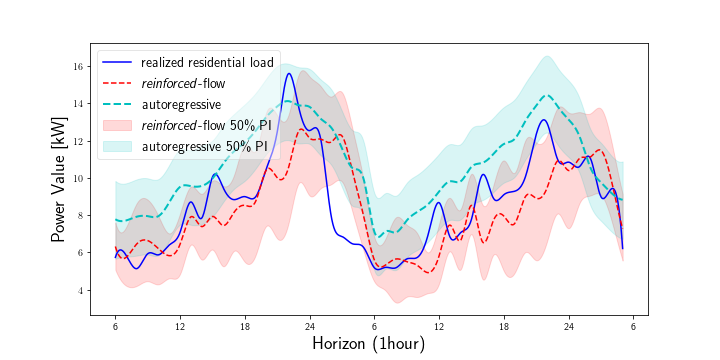}
\label{AR-reinforced}}
\caption{
Scenarios generated by \textit{vanilla}-flow (a) and \textit{reinforced}-flow (b) versus those by adding noise to point forecasts from the autoregressive model.
The blue curve is the realized residential load, dashed curves represent the medians of the generated scenarios, and colored banded areas indicate the $50\%$ prediction interval. In both structures, scenario generated by our approach are more accurate in forecasting the future load evolution and also more diverse than those from the standard method.
}
\label{comparison}
\end{figure}
In Fig. \ref{comparison}, we compare the generated scenario forecasts using our approach against those using the standard method of adding noise to point forecasts by plotting the medians and the $50\%$ prediction intervals of generated scenarios. Particularly, to get the point forecasts for the future 24 hours' residential load, we develop a 24-order linear autoregressive model to perform recursive multi-step forecasting~\cite{Marcellino2006MultistepForecasting}. In Fig. \ref{AR-vanilla}, we compare the scenario forecasts from \textit{vanilla}-flow with those from the standard method, and compare the forecasting results of \textit{reinforced}-flow with those of the standard method in Fig. \ref{AR-reinforced}. We can see from Fig. \ref{comparison} that our flow-based approach outperforms the standard method by providing more accurate forecasts for the future load in the form of median forecasts under both structure designs. Besides, the $50\%$ prediction interval of the generated scenarios using our approach can be more reliable by surrounding the realized load closely. In contrast, the scenarios generated by the standard method deviate from the realized load and fail to accurately forecast the coming time and power value for the future load valley. It is also worth pointing out that, compared to the unchanged width of the $50\%$ prediction interval from the standard method, the scenarios generated by our approach have smaller variance at load valley and larger variance at load peak to show different confidence levels about the forecasting results.

\subsection{Quantitive Evaluation}
\label{quantitive-evaluation}
In the forecasting literature, two properties have been used to examine the quality of generated scenario forecasts: reliability and sharpness~\cite{Murphy1993GoodForecast, Pinson2006PFQuality}. The reliability of generated scenario forecasts is that they should cover the actual value as much as possible, and the sharpness requires that the generated scenario forecasts should be able to provide a situation-dependent assessment of the forecast uncertainty. For the example of residential load scenario forecasting, it is intuitively expected that the forecast uncertainty should not be the same when the power consumption drops to the lowest point and when the peak demand occurs, because the smallest level of load demand is usually the base load and can be more fixed than the peak demand.

To examine the quality of the generated scenarios using our approach, we first evaluate the reliability of the generated scenarios by using the ``Deviation-Coverage Area" plot. Particularly, the Deviation-Coverage Area plot gives the amount of deviation from the realized load as a function of the size in the coverage area of quantiles.
Specifically, considering a coverage area of size $1-\alpha$, $0\leq \alpha \leq 1$, then we calculate the $\alpha/2^{\mbox{th}}$ and $(1-\alpha/2)^{\mbox{th}}$ quantiles of the generated scenario forecasts at each time point $t$. Denote the value of realized residential load at time $t$ by $y_t$, the $\alpha/2^{\mbox{th}}$ and $(1-\alpha/2)^{\mbox{th}}$ quantiles of the scenario forecasts at time $t$ by $\hat{y}_t^{\alpha/2}$ and $\hat{y}_t^{1-\alpha/2}$, respectively. Then the amount of deviation from $y_t$ for a coverage area with size $1-\alpha$ can be calculated as follows:
\begin{align}
{dev}_t^{1-\alpha} = \left\{ \begin{array}{ll}
         0 & \mbox{if $y_t \in [\hat{y}_t^{\alpha/2}, \hat{y}_t^{1-\alpha/2}]$};\\
         \hat{y}_t^{\alpha/2} - y_t& \mbox{if $y_t < \hat{y}_t^{\alpha/2}$};\\
        y_t - \hat{y}_t^{1-\alpha/2} & \mbox{if $y_t > \hat{y}_t^{1-\alpha/2}$}.\end{array} \right.
\end{align}
Particularly, when $\alpha = 1$, the coverage area has size zero, and both the $\alpha/2^{\mbox{th}}$ and $(1-\alpha/2)^{\mbox{th}}$ quantiles are calculated to be the median. In this case, the deviation amount for the coverage area with size zero is simply the distance between the realized residential load and the median.
With the scenario forecasts over $T$ look-ahead time points generated, for the coverage area with size $1-\alpha$, we average the deviation amounts from the realized load values  over all $T$ time points:
\begin{align}
{Dev}^{1-\alpha} = \frac{1}{T}\sum_{t=1}^{T}{dev}_t^{1-\alpha}.
\end{align}
Intuitively, the closer the Deviation-Coverage Are plot is to the origin, the higher reliability.

\begin{figure}[ht]
\centering
\subfloat[Single household]{\includegraphics[width=3in]{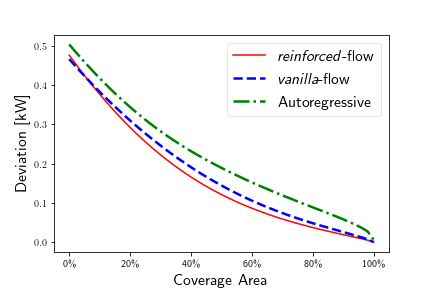}
\label{dev-area-level-1}}
\hfil
\subfloat[10 households]{\includegraphics[width=3in]{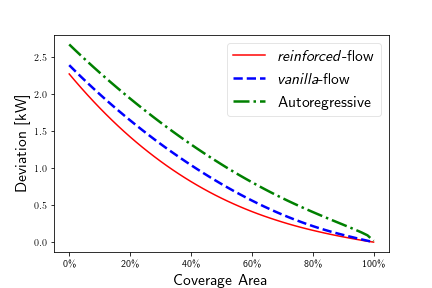}
\label{dev-area-level-10}}
\hfil
\subfloat[100 households]{\includegraphics[width=3in]{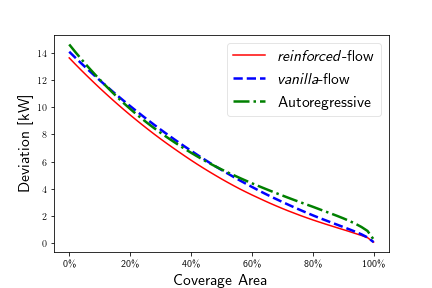}
\label{dev-area-level-100}}
\caption{
Deviation-Coverage Area plots of generated scenario forecasts for residential load of single household (a), 10 households (b), and 100 households (c).
Red curves are the Deviation-Coverage Area plots for the scenarios generated by \textit{reninforced}-flow, blue dashed curves are for \textit{vanilla}-flow, and green dotted curves are for the scenarios generated by adding noise to point forecasts. The generated scenarios using our approach have higher reliability than those coming from adding noise to point forecasts in all three aggregation settings, and \textit{reninforced}-flow shows a even better performance than \textit{vanilla}-flow.
}
\label{dev-area}
\end{figure}
The Deviation-Coverage Area plots of the generated scenarios for the residential load from a varying number of households are given in Fig. \ref{dev-area}. Particularly, we plot the Deviation-Coverage Area plots for both our approach and the standard method of adding noise to point forecasts for comparison.
We can see from Fig. \ref{dev-area} that our approach outperforms the standard scenario forecasting method of adding noise to point forecasts in all three aggregation levels, although the discrepancy between these two approaches decreases when 100 households are included. For the two structure designs used in our approach, the Deviation-Coverage Area plots for \textit{reinforced}-flow is closer to the origin in all three settings in Fig. \ref{dev-area}, and, therefore, the scenario forecasts generated by \textit{reinforced}-flow are more reliable than those by \textit{vanilla}-flow, which is also validated by the simulation results in Section \ref{qualitative-evaluation} in terms of the coverage of realized load using $50\%$ prediction interval.

% An example of a double column floating figure using two subfigures.
% (The subfig.sty package must be loaded for this to work.)
% The subfigure \label commands are set within each subfloat command,
% and the \label for the overall figure must come after \caption.
% \hfil is used as a separator to get equal spacing.
% Watch out that the combined width of all the subfigures on a
% line do not exceed the text width or a line break will occur.
%

\begin{figure}[ht]
\centering
\subfloat[Single household]{\includegraphics[width=3in]{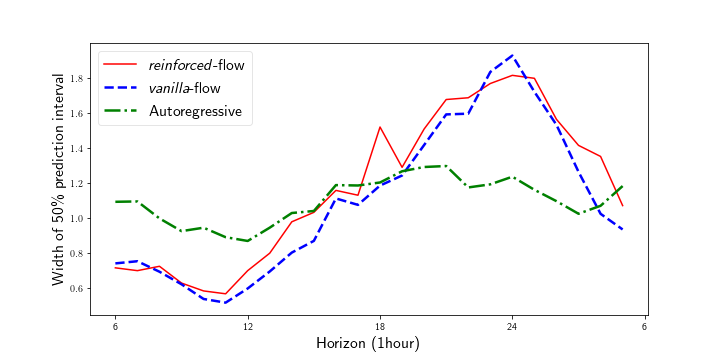}
\label{sharpness-1}}
\hfil
\subfloat[10 households]{\includegraphics[width=3in]{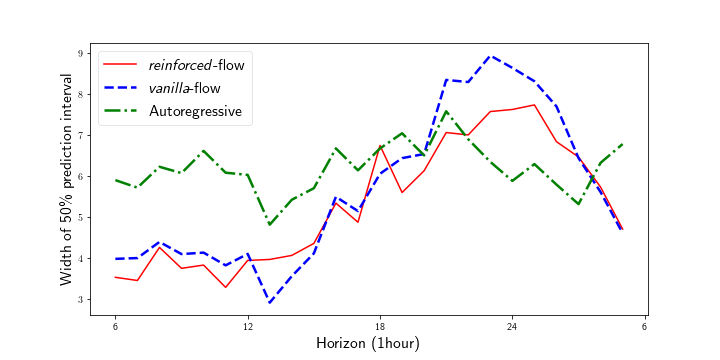}
\label{sharpness-10}}
\hfil
\subfloat[100 households]{\includegraphics[width=3in]{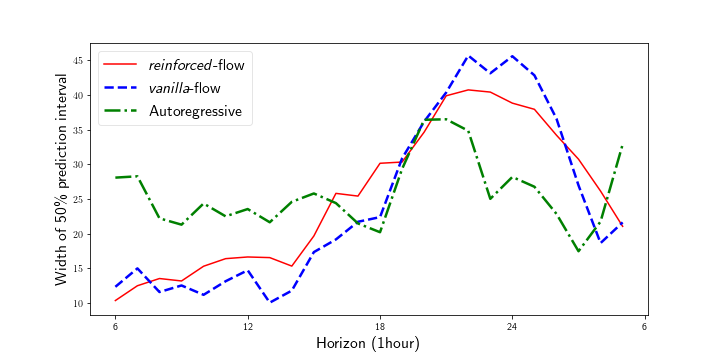}
\label{sharpness-100}}
\caption{The variations of the $50\%$ prediction interval width during one day for the generated scenario forecasts when single household (a), 10 households (b), and 100 households (c) are included.
The x-axis is the forecasting horizon at an hourly resolution, and the y-axis is the width of the $50\%$ prediction interval of generated scenarios in the unit of kW.
%The $50\%$ prediction interval width for the scenario forecasts generated using \textit{reinforced}-flow are given by red curves, for those using \textit{vanilla}-flow are plotted as blue dashed curves, and for those using the standard method of adding noise to point forecasts are shown as green dotted lines.
Compared to using the standard method (green dotted lines), the prediction interval width calculated from the scenarios generated by our approach using the two structure designs, \textit{reinforced}-flow (red lines) and \textit{vanilla}-flow (blue dashed lines), has more marked variations.
}
\label{sharpness}
\end{figure}
To evaluate the sharpness of the generated scenario forecasts, we calculate the width of the $50\%$ prediction interval, and investigate its variations during one day. The width of the $50\%$ prediction interval for the scenarios generated by our approach using two structure designs is given in Fig. \ref{sharpness}, and is compared against that from the standard method of adding noise to point forecasts. We can see from Fig. \ref{sharpness} that the $50\%$ prediction interval of the scenarios generated using our approach has pronounced varying width over 24 hours in all three aggregation settings. Particularly, if compared to the realized load in Section \ref{qualitative-evaluation}, the width can be narrower when the load is relatively small to show more confidence in the forecasting results, and becomes larger when the load demand reaches the high point in order for the $50\%$ prediction interval to cover a wider range of possibilities. By contrast, the variations in the prediction interval width of the scenarios generated by the standard method is not as noticeable as those in our approach.

%It can be seen from Fig. \ref{sharpness} that both the \textit{reinforced}-flow and the \textit{vanilla}-flow show obvious patterns in the $50\%$ prediction interval width of their produced scenarios at all three aggregation levels. Particularly, the \textit{vanilla}-flow has a relatively larger range of variations in the interval size than the \textit{reinforced}-flow.
%By comparing with Fig. \ref{all-levels}, we can see, in both proposed models, the largest interval size occurs at peak load, and the smallest occurs at the peak valley. \todo{Since the load peak is much more unpredictable than the load valley due to changes, it is expected that larger forecast uncertainty should be considered around this point.}
%In contrast, the benchmark approach of adding noise to point forecasts has minimal variations in the interval size at the aggregation level 1 and level 10. Although it shows a relatively more obvious pattern in the changes when aggregating 100 households for scenario forecasting, the benchmark method still has the smallest range of variations among all three methods.

\subsection{Cases Where Our Methods May Fail}
\begin{figure*}[ht]
\centering
\includegraphics[width=5in]{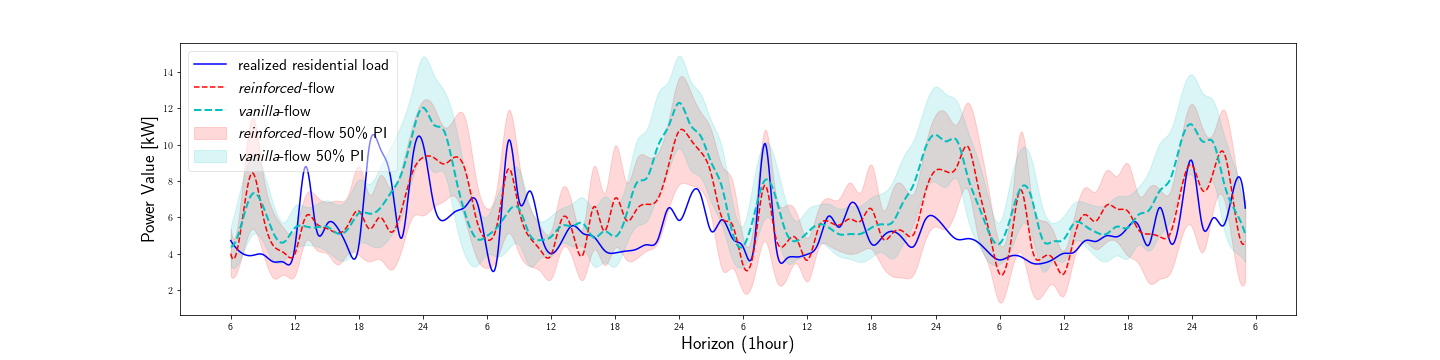}
\caption{A case where the scenarios generated by our approach using the two structure designs, \textit{reinforced}-flow (red) and \textit{vanilla}-flow (cyan), may fail to provide useful forecasts. The blue curve is the realized residential load, the dashed curves represent the medians of the generated scenarios, and the colored banded areas indicate the $50\%$ prediction interval.
Starting from the second day, the realized residential load becomes too noise to be learned useful information from. As a result, the flow-based model used in our approach ignores the noisy historical observations and generate scenarios that are irrelevant to the condition.
}
\label{fail}
\end{figure*}
In the simulation experiments, we also find there are certain cases where the generated scenarios using our approach may fail to provide accurate forecasts for the future load, and the derived prediction interval may not be able to cover the realized data. To show this, we give an example in Fig. \ref{fail}. We can see from this figure, the realized load becomes noisy and remains at a small level in most time. Although one can hardly
detect a clear pattern from the realized load in Fig. \ref{fail}, the median of the generated scenarios using our approach shows a clear pattern that is different than the realized load, and the $50\%$ prediction interval surrounds the median closely so as to not cover the realized load.
%may ignore the past information and fail to follow the actual load values. To show this, we give an example from the testing stage at aggregation level 10 in Fig. \ref{fail}. As we can see, starting from the second day, the realized residential load becomes very disordered and keeps relatively small values in most time. Although one can hardly detect a clear pattern from the realized residential load in Fig. \ref{fail}, both the median and the $50\%$ prediction interval of the scenarios coming from our methods present a normal residential load pattern on the third and fourth days, with peaks at 24:00 and valleys at some point between 06:00 and 12:00, which is nothing like the realized residential load curve.
The inaccurate forecasts of the generated scenarios in this case
%This failure to keep track of the past information
can be explained from the perspective of the objective function used for training flow-based models. Recall from Section \ref{conditional-flows} that the objective of training flow-based conditional generative models is to maximize the conditional log-likelihood on the training set. That it to say, given the historical observations, a flow-based model always generates the scenarios that are most likely to occur under this condition.
However, if the historical observations to be conditioned on is too noisy, then they can not provide useful side information to the flow-based model. Without the help of additional information, the flow-based model simply maximizes the log-likelihood instead of conditional log-likelihood for this training sample. As a result, the scenarios that are likely to occur under normal circumstances are generated by the flow-based model.

\subsection{Conditional Scenario Generation with Wasserstein Distance}
Recall from Section \ref{W-flows} that we add a weighted Wasserstein distance metric to the training objective of flow-based generative models to balance the large variety of scenarios generated by minimizing the KL divergence and the small variance of those generated by minimizing the Wasserstein distance. In order to validate if the scenarios generated by minimizing both metrics can have smaller variance compared to those generated by only minimizing the KL divergence metric, we use the residential load of 10 households, and show the conditional scenario forecasting results in both cases. Considering the better performance of \textit{reinforced}-flow than \textit{vanilla}-flow as shown in previous simulation results, we take the \textit{reinforced}-flow and its counterpart \textit{reinforced}-$\mathcal{W}$-flow as examples to illustrate the effect of adding a weighted Wasserstein distance metric.

Particularly, the Wasserstein distance between the true data distribution and the learned one is calculated through the dual formulation~\cite{villani2008optimal}:
\begin{align}
W(p_X(\mathbf{x}|\theta^{\star}), p_X(\mathbf{x}|\theta)) & = \sup_{g:\|g\|_L\leq 1} \mathbb{E}_{\mathbf{x} \sim p_X(\mathbf{x}|\theta^{\star})}[g(\mathbf{x})]\\
& - \mathbb{E}_{\mathbf{x} \sim p_X(\mathbf{x}|\theta)}[g(\mathbf{x})]
\label{Wasserstein-dual}
\end{align}
where $g$ represents any $1$-Lipschitz function that maps from $\mathbb{R}^D$ to $\mathbb{R}$, and $D$ is the dimension of $\mathbf{x}$. In our case of conditional scenario generation, the $1$-Lipschitz function $g$ is implemented as an 1D-CNN with the condition as an extra input. To enforce the $1$-Lipschitz property on $g$, we utilize the weight clamping technique in \cite{Arjovsky2017WGAN}.

\begin{figure}[ht]
\centering
\subfloat[A good case of realized residential load]{\includegraphics[width=3in]{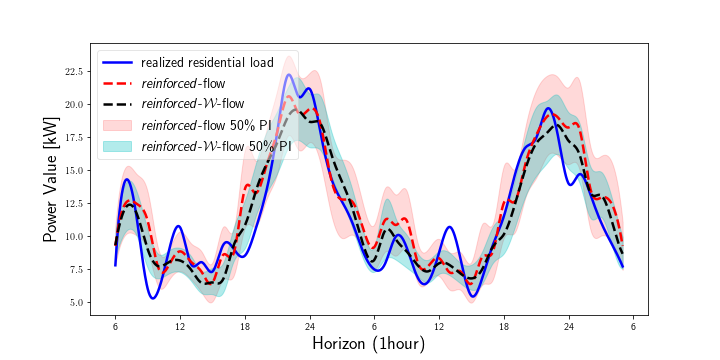}
\label{reinforced-W-good}}
\hfil
\subfloat[A bad case of realized residential load]{\includegraphics[width=3in]{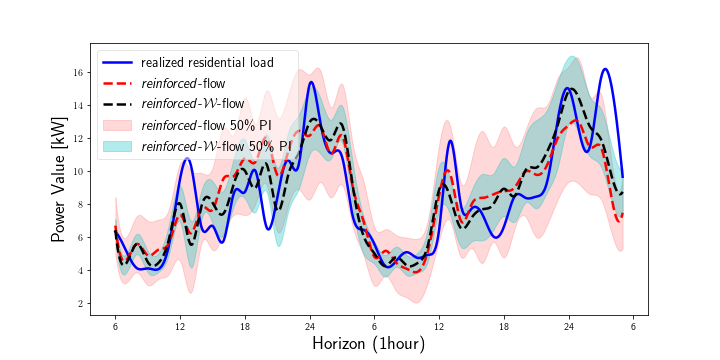}
\label{reinforced-W-bad}}
\caption{
Two examples of generated scenarios by minimizing both KL divergence and Wasserstein distance (\textit{reinforced}-$\mathcal{W}$-flow) versus only minimizing the KL divergence (\textit{reinforced}-flow).
Example (a) represents the case where the realized residential load has less volatility, while (b) represents the case where the realized residential load has large randomness.
The dashed lines represent the medians of the generated scenarios, and the colored banded areas indicate the $50\%$ prediction intervals.
In both examples, the scenarios generated by minimizing both metrics have noticeably narrower $50\%$ prediction intervals.}
\label{fig:W-flows-sample}
\end{figure}
In Fig. \ref{fig:W-flows-sample}, we show the scenarios generated by the \textit{reinforced}-flow and its counterpart \textit{reinforced}-$\mathcal{W}$-flow, and plot the medians and the $50\%$ prediction intervals of the scenarios generated by these two models.
Particularly, to validate that we can reduce the variance of the generated scenarios by adding a weighted Wasserstein distance in both cases of large uncertainties and small uncertainties,
we have shown two samples in Fig. \ref{fig:W-flows-sample}, where the residential load in sample Fig. \ref{reinforced-W-bad} has larger volatility than that in sample Fig. \ref{reinforced-W-bad}.
We can see from Fig. \ref{fig:W-flows-sample} that, the scenarios generated by \textit{reinforced}-$\mathcal{W}$-flow have smaller variance at all time points in both samples compared to those generated by \textit{reinforced}-flow. Particularly, the reduction in variance is relatively larger in sample (b) when the realized load curve is more volatile compared to that in sample (a).
It is also worth pointing out that, in each sample, the scenarios generated by \textit{reinforced}-$\mathcal{W}$-flow can have even smaller variance than those generated by \textit{reinforced}-flow when the residential load to be forecasted is more certain, i.e., at load valley. When the residential load becomes more unpredictable, i.e., at load peak, the $50\%$ prediction interval of the scenarios generated by \textit{reinforced}-$\mathcal{W}$-flow also becomes wider but is still narrower than that in \textit{reinforced}-flow. We also note that, because of the relatively small variance, the median of the scenarios generated by \textit{reinforced}-$\mathcal{W}$-flow can provide more accurate forecasts for the future residential load evolution than those generated by \textit{reinforced}-flow.

\begin{figure}[ht]
\centering
\includegraphics[width=3in]{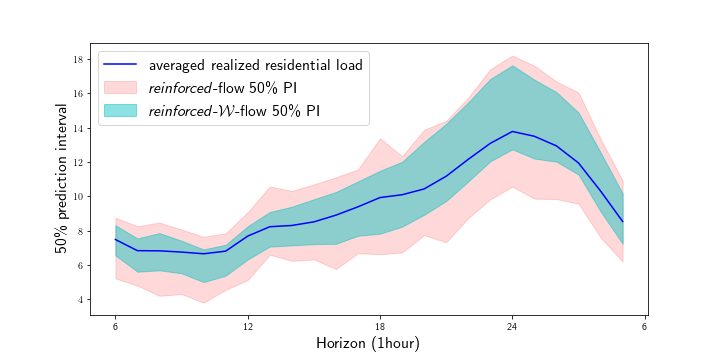}
\caption{
The $50\%$ prediction interval of scenarios generated by \textit{reinforced}-$\mathcal{W}$-flow (cyan colored band) versus those generated by \textit{reinforced}-flow (red colored band). The blue curve is the averaged daily realized load. The scenarios generated by \textit{reinforced}-$\mathcal{W}$-flow have narrower $50\%$ prediction interval at all time points compared to those generated by \textit{reinforced}-flow.
}
\label{fig:W-flow-sharpness}
\end{figure}
To quantitatively investigate the effect of adding a weighted Wasserstein distance metric, we calculate the $50\%$ prediction intervals of all generated daily scenarios and plot the averaged daily interval in Fig. \ref{fig:W-flow-sharpness}. As can be seen from Fig. \ref{fig:W-flow-sharpness}, the realized load is surrounded by the $50\%$ prediction intervals in both models. However, the $50\%$ prediction interval of the scenarios generated by \textit{reinforced}-$\mathcal{W}$-flow has narrower width at all time points in one day compared to those generated by \textit{reinforced}-flow. That is to say, the scenarios generated by minimizing both KL divergence and Wasserstein distance metrics can have smaller variance than those generated by only minimizing KL divergence. It is also worth noting that, similar to the behavior of the generated scenarios using \textit{reinforced}-flow, the scenarios generated by \textit{reinforced}-$\mathcal{W}$-flow also show narrower $50\%$ prediction interval at the load valley than at the load peak. That is to say, by adding a weighted Wasserstein distance, we can reduce the variance of generated scenarios but still maintain original sharpness.

\section{Conclusion}
\label{conclusions}
In this paper we study the problem of scenario forecasting for a single or a small group of households. Because of the advancement of solar PV, electric vehicles and demand response, accurately forecasting the behavior at the household level becomes an important step in the operation of distribution systems. Since the high variability and small base load make providing accurate point forecasts difficult, we focus on providing a group of scenarios that capture the potential behavior of future load. We adopt the so-called flow-based generation method, where we generate time series representing the future load conditioned on past historical observations. This approach can generate scenarios that are both diverse and follows the true pattern of the load.
The simulation results show the flow-based designs significantly outperform existing methods in scenario forecasting for residential load by providing both more accurate and more diverse scenarios.

\bibliographystyle{IEEEtran}
\bibliography{mybib}

\begin{thebibliography}{10}
\providecommand{\url}[1]{#1}
\csname url@samestyle\endcsname
\providecommand{\newblock}{\relax}
\providecommand{\bibinfo}[2]{#2}
\providecommand{\BIBentrySTDinterwordspacing}{\spaceskip=0pt\relax}
\providecommand{\BIBentryALTinterwordstretchfactor}{4}
\providecommand{\BIBentryALTinterwordspacing}{\spaceskip=\fontdimen2\font plus
\BIBentryALTinterwordstretchfactor\fontdimen3\font minus
  \fontdimen4\font\relax}
\providecommand{\BIBforeignlanguage}[2]{{%
\expandafter\ifx\csname l@#1\endcsname\relax
\typeout{** WARNING: IEEEtran.bst: No hyphenation pattern has been}%
\typeout{** loaded for the language `#1'. Using the pattern for}%
\typeout{** the default language instead.}%
\else
\language=\csname l@#1\endcsname
\fi
#2}}
\providecommand{\BIBdecl}{\relax}
\BIBdecl

\bibitem{Guo2012ManagePower}
Y.~Guo, M.~Pan, and Y.~Fang, ``Optimal power management of residential
  customers in the smart grid,'' \emph{Parallel and Distributed Systems, IEEE
  Transactions on}, vol.~23, pp. 1593--1606, 2012.

\bibitem{Ji2018DataDrivenLM}
Y.~Ji, E.~D. {Buechler}, and R.~Rajagopal, ``Data-driven load modeling and
  forecasting of residential appliances,'' 2018.

\bibitem{Tascikaraoglu2014DSM}
A.~Taşcıkaraoğlu, A.~Boynuegri, and M.~Uzunoglu, ``A demand side management
  strategy based on forecasting of residential renewable sources: A smart home
  system in turkey,'' \emph{Energy and Buildings}, vol.~80, pp. 309--320, 2014.

\bibitem{Sevlian2018VaryLevels}
R.~Sevlian and R.~Rajagopal, ``A scaling law for short term load forecasting on
  varying levels of aggregation,'' \emph{International Journal of Electrical
  Power \& Energy Systems}, vol.~98, pp. 350--361, 2018.

\bibitem{veit2014household}
A.~Veit, C.~Goebel, R.~Tidke, C.~Doblander, and H.-A. Jacobsen, ``Household
  electricity demand forecasting--benchmarking state-of-the-art methods,''
  \emph{arXiv preprint arXiv:1404.0200}, 2014.

\bibitem{Humeau13}
S.~{Humeau}, T.~K. {Wijaya}, M.~{Vasirani}, and K.~{Aberer}, ``Electricity load
  forecasting for residential customers: Exploiting aggregation and correlation
  between households,'' in \emph{2013 Sustainable Internet and ICT for
  Sustainability (SustainIT)}, Oct 2013, pp. 1--6.

\bibitem{Gajowniczek2017IndvForecast}
K.~Gajowniczek and T.~Ząbkowski, ``Electricity forecasting on the individual
  household level enhanced based on activity patterns,'' \emph{PLoS ONE},
  vol.~12, 2017.

\bibitem{Zhang2018SinglePerspect}
X.~Monica~Zhang, K.~Grolinger, M.~Capretz, and L.~Seewald, ``Forecasting
  residential energy consumption: Single household perspective,'' in \emph{IEEE
  International Conference on Machine Learning and applications}, 2018, pp.
  110--117.

\bibitem{Yildiz2018IndvForecast}
B.~Yildiz, J.~Bilbao, J.~Dore, and A.~Sproul, ``Short-term forecasting of
  individual household electricity loads with investigating impact of data
  resolution and forecast horizon,'' \emph{Renewable Energy and Environmental
  Sustainability}, vol.~3, p.~3, 2018.

\bibitem{Hong2016Tutorial}
T.~Hong and S.~Fan, ``Probabilistic electric load forecasting: A tutorial
  review,'' \emph{International Journal of Forecasting}, vol.~32, no.~3, pp.
  914–--938, 2016.

\bibitem{Hyndman2014Forecasting}
R.~J. Hyndman and G.~Athanasopoulos, \emph{Forecasting: principles and
  practice}.\hskip 1em plus 0.5em minus 0.4em\relax OTexts, 2014.

\bibitem{Hippert2001NN}
H.~S. Hippert, C.~Pedreira, and R.~Castro~Souza, ``Neural networks for
  short-term load forecasting: A review and evaluation,'' \emph{Power Systems,
  IEEE Transactions on}, vol.~16, pp. 44--55, 2001.

\bibitem{Chen2004Competition}
B.-J. Chen, M.-W. Chang, and C.-J. Lin, ``Load forecasting using support vector
  machines: A study on eunite competition 2001,'' \emph{Power Systems, IEEE
  Transactions on}, vol.~19, pp. 1821--1830, 2004.

\bibitem{Weron2006StatForecast}
R.~Weron, \emph{Modeling and Forecasting Electricity Loads and Prices: A
  Statistical Approach}, 01 1998.

\bibitem{Pflugradt2017SynthLoad}
N.~Pflugradt and U.~Muntwyler, ``Synthesizing residential load profiles using
  behavior simulation,'' \emph{Energy Procedia}, vol. 122, pp. 655--660, 2017.

\bibitem{Narayan2018ConstructLoad}
N.~Narayan, Z.~Qin, J.~Popovic-Gerber, J.-C. Diehl, P.~Bauer, and M.~Zeman,
  ``Stochastic load profile construction for the multi-tier framework for
  household electricity access using off-grid dc appliances,'' \emph{Energy
  Efficiency}, 2018.

\bibitem{Linssen2017LoadInfluence}
``Techno-economic analysis of photovoltaic battery systems and the influence of
  different consumer load profiles,'' \emph{Applied Energy}, vol. 185, pp.
  2019--2025, 2017, clean, Efficient and Affordable Energy for a Sustainable
  Future.

\bibitem{Xie2017NormalAssump}
J.~{Xie}, T.~{Hong}, T.~{Laing}, and C.~{Kang}, ``On normality assumption in
  residual simulation for probabilistic load forecasting,'' \emph{IEEE
  Transactions on Smart Grid}, vol.~8, no.~3, pp. 1046--1053, 2017.

\bibitem{McSharry2005PeakPLF}
P.~E. {McSharry}, S.~{Bouwman}, and G.~{Bloemhof}, ``Probabilistic forecasts of
  the magnitude and timing of peak electricity demand,'' \emph{IEEE
  Transactions on Power Systems}, vol.~20, no.~2, pp. 1166--1172, 2005.

\bibitem{Fan2012ShortTerm}
S.~{Fan} and R.~J. {Hyndman}, ``Short-term load forecasting based on a
  semi-parametric additive model,'' \emph{IEEE Transactions on Power Systems},
  vol.~27, no.~1, pp. 134--141, 2012.

\bibitem{Dordonnat2016GEFCom}
V.~Dordonnat, A.~Pichavant, and A.~Pierrot, ``Gefcom2014 probabilistic electric
  load forecasting using time series and semi-parametric regression models,''
  \emph{International Journal of Forecasting}, vol.~32, no.~3, pp. 1005--1011,
  2016.

\bibitem{Hong2018PLF}
J.~{Xie} and T.~{Hong}, ``Temperature scenario generation for probabilistic
  load forecasting,'' \emph{IEEE Transactions on Smart Grid}, vol.~9, no.~3,
  pp. 1680--1687, 2018.

\bibitem{Chen2018ModelFreeRS}
Y.~Chen, Y.~Wang, D.~S. Kirschen, and B.~Zhang, ``Model-free renewable scenario
  generation using generative adversarial networks,'' \emph{IEEE Transactions
  on Power Systems}, vol.~33, pp. 3265--3275, 2018.

\bibitem{Chen2018Unsupervised}
Y.~Chen, X.~Wang, and B.~Zhang, ``An unsupervised deep learning approach for
  scenario forecasts,'' \emph{2018 Power Systems Computation Conference
  (PSCC)}, pp. 1--7, 2018.

\bibitem{Goodfellow2014GenerativeAN}
I.~J. Goodfellow, J.~{Pouget-Abadie}, M.~Mirza, B.~Xu, D.~{Warde-Farley},
  S.~Ozair, A.~C. {Courville}, and Y.~Bengio, ``Generative adversarial nets,''
  in \emph{Advances in Neural Information Processing Systems}, 2014, pp.
  2672--2680.

\bibitem{Gu2018GANbasedRL}
Y.~Gu, Q.~Chen, K.~Liu, L.~Xie, and C.~Kang, ``Gan-based model for residential
  load generation considering typical consumption patterns,'' in
  \emph{Conference on Innovative Smart Grid Technologies}, 11 2018.

\bibitem{Odena2017ConditionalIS}
A.~Odena, C.~Olah, and J.~Shlens, ``Conditional image synthesis with auxiliary
  classifier gans,'' in \emph{International Conference on Machine Learning},
  2017.

\bibitem{Mirza2014ConditionalGAN}
M.~Mirza and S.~Osindero, ``Conditional generative adversarial nets,''
  \emph{CoRR}, vol. abs/1411.1784, 2014.

\bibitem{Arjovsky2017WGAN}
M.~Arjovsky, S.~Chintala, and L.~Bottou, ``Wasserstein generative adversarial
  networks,'' in \emph{International Conference on Machine Learning}, 2017, pp.
  214--223.

\bibitem{Dinh2014NICE}
L.~Dinh, D.~Krueger, and Y.~Bengio, ``Nice: Non-linear independent components
  estimation,'' \emph{CoRR}, vol. abs/1410.8516, 2014.

\bibitem{Dinh2017RealNVP}
L.~Dinh, J.~{Sohl-Dickstein}, and S.~{Bengio}, ``Density estimation using real
  nvp,'' \emph{CoRR}, vol. abs/1605.08803, 2017.

\bibitem{Kingma2018GlowGF}
D.~P. Kingma and P.~Dhariwal, ``Glow: Generative flow with invertible 1x1
  convolutions,'' in \emph{Advances in Neural Information Processing Systems},
  2018, pp. 10\,236–--10\,245.

\bibitem{Kingma2014AutoEncodingVB}
D.~P. Kingma and M.~Welling, ``Auto-encoding variational bayes,'' \emph{CoRR},
  vol. abs/1312.6114, 2014.

\bibitem{Murphy2012MLP}
K.~P. Murphy, \emph{Machine Learning: A Probabilistic Perspective}.

\bibitem{durrett2019probability}
R.~Durrett, \emph{Forecasting: Principles and Practice}.\hskip 1em plus 0.5em
  minus 0.4em\relax Cambridge university press, 2019, vol.~49.

\bibitem{ruschendorf1985wasserstein}
L.~R{\"u}schendorf, ``The wasserstein distance and approximation theorems,''
  \emph{Probability Theory and Related Fields}, vol.~70, no.~1, pp. 117--129,
  1985.

\bibitem{Paszke2017AutomaticDI}
A.~Paszke, S.~Gross, S.~Chintala, G.~Chanan, E.~Yang, Z.~DeVito, Z.~Lin,
  A.~Desmaison, L.~Antiga, and A.~Lerer, ``Automatic differentiation in
  pytorch,'' in \emph{Conference on Neural Information Processing Systems},
  2017.

\bibitem{Kingma2015Adam}
D.~P. {Kingma} and J.~Ba, ``Adam: A method for stochastic optimization,''
  \emph{CoRR}, vol. abs/1412.6980, 2015.

\bibitem{web:pecanstreet}
\BIBentryALTinterwordspacing
P.~S. Inc. [Online]. Available: \url{https://dataport.pecanstreet.org}
\BIBentrySTDinterwordspacing

\bibitem{Marcellino2006MultistepForecasting}
M.~Marcellino, J.~Stock, and M.~W.~Watson, ``A comparison of direct and
  iterated multistep ar methods for forecasting macroeconomic time series,''
  \emph{Journal of Econometrics}, vol. 135, pp. 499--526, 2006.

\bibitem{Murphy1993GoodForecast}
A.~H. Murphy, ``What is a good forecast? an essay on the nature of goodness in
  weather forecasting,'' \emph{Weather and Forecasting}, vol.~8, no.~2, pp.
  281--293, 1993.

\bibitem{Pinson2006PFQuality}
P.~{Pinson}, J.~{Juban}, and G.~N. {Kariniotakis}, ``On the quality and value
  of probabilistic forecasts of wind generation,'' in \emph{2006 International
  Conference on Probabilistic Methods Applied to Power Systems}, 2006, pp.
  1--7.

\bibitem{villani2008optimal}
\BIBentryALTinterwordspacing
C.~Villani, \emph{Optimal Transport: Old and New}, ser. Grundlehren der
  mathematischen Wissenschaften.\hskip 1em plus 0.5em minus 0.4em\relax
  Springer Berlin Heidelberg, 2008. [Online]. Available:
  \url{https://books.google.com/books?id=hV8o5R7\_5tkC}
\BIBentrySTDinterwordspacing

\end{thebibliography}

\end{document}